\documentclass[aps,pre,superscriptaddress,longbibliography]{revtex4-2}

\usepackage{amsmath}
\usepackage{amssymb}
\usepackage{graphicx}
\usepackage{hyperref}
\usepackage{xcolor}
\usepackage{booktabs}
\usepackage{dcolumn}    
\usepackage{multirow}   

\newcommand{\paperone}{\cite{BernalAlvarado2026a}}
\newcommand{\orbis}{\textsc{orbis}}

\newcommand{\Hstar}{H^*}

\begin{document}

\title{Structural Divergence of the Roman--Byzantine Trade Network,
0--1453\,CE: Persistent Homology, Topological Velocity,
and Criticality Indicators of Imperial Collapse}

\author{José de Jesús Bernal-Alvarado}
\email{bernal@ugto.mx}
\affiliation{Physics Engineering Department,
       Universidad de Guanajuato, México}

\author{David Delepine}
\email{delepine@ugto.mx}
\affiliation{Physics Department,
       Universidad de Guanajuato, México}

\author{Carlos Pinedo Guadarrama}
\email{c.pinedoguadarrama@ugto.mx}
\affiliation{Physics Department,
       Universidad de Guanajuato, México}

\date{\today}

\begin{abstract}

We extend the persistent homology analysis of~\paperone{}
to the full Roman--Byzantine trade network
(0--1453\,\textsc{ce}), using 2{,}599 nodes and 4{,}503
trimodal edges calibrated against the \textsc{orbis}
Geospatial Network Model.
Five results are reported.
(i)~The $H_t{=}0$ western sub-network result of~\paperone{}
is a data-coverage artifact: with full western representation
($N_{\rm west}=987$, $\beta_1\approx52$ cycles per decade)
a baseline East--West entropy gap of $+2.22$ units is present
from 0\,\textsc{ce} and grows at $+3.3\times10^{-3}$\,yr$^{-1}$,
predating the Theodosian partition by four centuries.
(ii)~A \emph{hub-selection artifact} in degree-heterogeneous
networks can reverse the sign of the inferred Phase~III slope,
requiring full-coverage or stratified sampling for reliable
structural-break detection.
(iii)~Decomposing Byzantine resilience into geographic
($H_{\rm geo}$) and economic ($H_{\rm eco}$) components
reveals a peak decoupling ratio
$R_d = H_{\rm eco}/H_{\rm geo} = 47.7$ at 620\,\textsc{ce},
falling to 13.9 at 640\,\textsc{ce}, quantifying the
McCormick--Ward-Perkins historiographical debate as a
contrast between two network layers operating on different
timescales.
(iv)~The inter-decade $W_2$ Wasserstein velocity identifies
the Late Roman--Early Byzantine transition
(495\,\textsc{ce}) as the highest topological-velocity event
of the 1,453-year record; the cross-network Wasserstein
ratio increases by $150$--$300\times$ after the Chrysobull
of 1082\,\textsc{ce}, providing an independent diagram-space
analogue of $R_d$.
Both the Western collapse (476\,\textsc{ce}) and the
Byzantine endpoint (1453\,\textsc{ce}) occur at
$H^{\ast}\approx0.524$, interpreted as a candidate
topological percolation threshold.
\end{abstract}

\maketitle

\section{Introduction}


The companion study to this work~\paperone{} demonstrated that the
Eastern Mediterranean trade network --- as represented in the
\textsc{orbis} Geospatial Network Model of the Roman
World~\cite{Meeks2013,Scheidel2014} and situated within
the broader economic history of the Greco-Roman
world~\cite{Scheidel2007} --- exhibited three
statistically distinct phases in its $\beta_1$ persistent entropy
time series between 0 and 400\,\textsc{ce}, with an irreversible
structural break at 310\,\textsc{ce} (Chow $F=85.4$,
$p<0.0001$) from which no topological recovery was observed
despite Constantinian political reunification.
The same study reported that the western sub-network produced
$H_t=0$ throughout the study period and acknowledged this as a
potential data limitation arising from the restricted western
coverage of the \textsc{orbis} extract employed: 49 nodes with
no representation of Italy, Gaul, Britannia, or the Rhine
frontier.

The present study provides a complete western dataset, integrating 987 western nodes
from the Pleiades Digital Atlas of the Ancient
World~\cite{Pleiades} with road geometry from the \textsc{darmc}
Roman Roads Network~\cite{DARMC}, maritime routes derived from
\textsc{orbis} coastal and overseas edge topology, and provincial
boundary polygons~\cite{Romanprov} for temporal node activation.

The western sub-network generated $\beta_1\approx52$ independent
cycles throughout 0--400\,\textsc{ce}, and exhibited measurable
entropy throughout that period, with a Phase~I mean of
$\bar{H}_{\rm west}=1.362$.
The $H_t=0$ finding of~\paperone{} characterised the
\textsc{orbis} extract, not the actual western network.
The structural question becomes not whether the West had
topological resilience, but how much less it had than the East
from the very beginning.


The East--West structural comparison generates three distinct
questions, each independently testable with the
integrated dataset constructed here.

\begin{itemize}
\item \textbf{The baseline East--West asymmetry question.}
If $H_{\rm west} < H_{\rm east}$ at 0\,\textsc{ce} then the East--West structural difference was
not produced by the Crisis of the Third Century, the military
reforms of Diocletian, or the Theodosian division.
It was a property of the network from the earliest phase of
the Principate. 
This baseline structural
asymmetry implies that Western fragility was already embedded
in the network, rather than being imposed only by later external
pressures \cite{Wickham2005,Heather2006}.

\item \textbf{The topology-precedes-politics question.}
The political division of the Empire under Theodosius is the conventional historical marker of
East--West separation.
If the difference between $H_{\rm east}$ and $H_{\rm west}$ entropy grew
continuously from 0\,\textsc{ce} 
, the political partition formalised a
structural reality that had been developing since the beginning. 

\item \textbf{The Byzantine continuity question.}
The Eastern Roman Empire survived the fifth-century collapse and
continued for 977 further years until 1453\,\textsc{ce}.
Was this survival a topologically continuous process --- the same
network degrading at the same rate from a fourfold higher
starting point --- or did the Byzantine period constitute a
qualitatively distinct network regime?
\end{itemize}

Three well-known historical claims are
testable through topological network analysis:

\begin{itemize}

\item \textbf{Ward-Perkins (2005) and McCormick (2001).}
Ward-Perkins~\cite{WardPerkins2005} argues that the fifth-century
Western collapse was a genuine catastrophe proved by the
decline of material culture indicators implying
a severe contraction of long-distance commercial exchange
circuits.
McCormick~\cite{McCormick2001} argues, based on a comprehensive
reconstruction of communications in the early medieval
Mediterranean, that the apparent collapse of Roman commerce
conceals a genuine continuity of exchange activity that adapted
to a new political situation.
The decomposition of Byzantine resilience into geographic
($H_{\rm geo}$) and economic ($H_{\rm eco}$) components provides
a metric capable of distinguishing and quantifying differences between these
two accounts: at 640\,\textsc{ce},
$H_{\rm eco}/H_{\rm geo}=13.9$, meaning the commercial
Mediterranean retained approximately fourteen times more
cycle redundancy than the imperial geographic infrastructure.
Ward-Perkins describes the geographic layer; McCormick describes
the economic one.

\item \textbf{Harper (2017).}
Harper~\cite{Harper2017} locates the primary drivers of Roman
imperial fragility in climate deterioration and pandemic
mortality.
The Justinianic Plague of 541\,\textsc{ce} produces the largest
Chow $F$-statistic in the 0--1453\,\textsc{ce} geographic series
(Chow $F=152.69$, $p<0.0001$), confirming that pandemic
disruption constituted a structural shock.
However, the plague signal is absent from the economic
network ($H_{\rm eco}$ unchanged at 541\,\textsc{ce}),
which is consistent with a demographic disruption rather than a permanent infrastructural disruption of the route system,
and therefore supports Harper's demographic argument. 

\item \textbf{Wickham (2005).}
Wickham~\cite{Wickham2005} argues that the transition from a
tax-and-redistribute imperial economy to localised subsistence
exchange is the structural transformation defining late
antiquity.
The baseline East--West structural asymmetry documented here ---
$+2.222$ entropy units in Phase~I --- supports Wickham's structural framing.
The Western network was not a Roman network that deteriorated
under external stress, but a structurally weaker commercial
system from its inception, whose fragility became decisive 
when the external conditions sustaining it were progressively withdrawn.
Wickham's structural transformation was not produced by the
Crisis of the Third Century: it was the eventual expression of
a structural difference that had existed since Augustus.

\end{itemize}


Three methods described in Ref.~\paperone{} are used 
here: the differential friction model assigning
transport-mode-specific volatility parameters to each edge; the
adaptive filtration threshold ($\delta_t = $ 90th percentile of
pairwise distances within the active sub-network at each decade);
and the Chow structural break test~\cite{Chow1960} for phase identification. For details, the reader is referred
to~\paperone{}, Section~II, for the complete specification.

The present study adds six methodological contributions.
Throughout the article, claims about collapse thresholds are treated as within-system hypotheses for the Roman--Byzantine network, not as established constants of pre-modern empires in general.

\begin{itemize}
  \item \textbf{Minimum-coverage convergence protocol.}
The hub-selection artifact --- a non-monotonic $N$-convergence
in which $N=800$ top-degree sampling produces lower Chow
$F$-statistics than both $N=400$ and $N=1{,}600$, because
degree-dominant nodes from one sub-network dilute the regional
signal of the other before both have reached sufficient sampling
density --- is documented here (Table~\ref{tab:convergence}).
The practical consequence is a minimum-coverage requirement:
structural break detection in degree-heterogeneous historical
networks requires either $N>N_{\min}$, where $N_{\min}$ is the
sampling threshold above which both sub-networks reach sufficient
density to reconstruct their independent cycles, or a sampling
protocol that explicitly preserves regional representation
ratios.

\item \textbf{Trimodal edge architecture.}
The \textsc{orbis} differential friction model is extended to
include terrestrial road edges (calibrated against \textsc{orbis}
anchor points via log-log regression, factor $8.90\times$),
maritime routes covering the western and eastern Mediterranean basin, and
five canal edges --- producing a network in which three transport
modes respond to documented historical
perturbations across the full geographic extent of the Empire.

\item \textbf{Geographic--economic network decomposition.}
For the Byzantine period (400--1453\,\textsc{ce}), two parallel
\textsc{tda} pipelines are computed: one based on
territorial control and the other on commercial relationships.
The structural decoupling between imperial territory and
commercial resilience is measured using the ratio $H_{\rm eco}/H_{\rm geo}$ in each decade.

\item \textbf{Macroeconomic Friction Index: Byzantine extension.}
The Macroeconomic Friction Index of~\paperone{} is extended over the entire study period using the McConnell
et~al.~\cite{McConnell2018} Greenland ice-core lead series as an
observed proxy for metallurgical activity (calibrated to monetary
production and trade volume) from $-50$ to 800\,\textsc{ce},
and literature-reconstructed control points from 810 to
1450\,\textsc{ce}, with explicit source-tier annotation
distinguishing observed, interpolated, and historically
calibrated data.

\item \textbf{Topological velocity via Wasserstein distances.}
The $W_2$ Wasserstein distance between consecutive decadal
persistence diagrams~\cite{Edelsbrunner2002,Kerber2017}, the Wasserstein or topological  velocity 
$\dot{W}_2(t) = W_2\bigl(\mathcal{D}_{t},\mathcal{D}_{t+10}\bigr)/10$, 
is computed for all four network layers and for all
cross-layer pairs throughout both the Roman (0--400\,\textsc{ce})
and Byzantine (400--1453\,\textsc{ce}) periods.

\item \textbf{Integrated Criticality Threshold (ICT).}
A composite early-warning indicator is constructed by normalising
and averaging three quantities derived from the persistence
diagrams at each decade: the topological susceptibility
$\chi(t)$ (variance of the normalised lifetime distribution),
the correlation length $\xi(t)$ (characteristic persistence
lifetime), and the normalised Wasserstein velocity
$\dot{W}_{2,\rm norm}(t)$:
\begin{equation}
\mathrm{ICT}(t) = \frac{\chi_{\rm norm}(t) + \xi_{\rm norm}(t)
                         + \dot{W}_{2,\rm norm}(t)}{3}.
\label{eq:ICT}
\end{equation}
An $\mathrm{ICT}$ value approaching 1 indicates that the network
is operating in a regime analogous to the critical point of a
statistical-mechanical phase transition: maximum susceptibility
to perturbations and long-range correlation of topological
features.
\end{itemize}

Section~\ref{sec:data} describes the integrated dataset, the
trimodal edge architecture, the Macroeconomic Friction Index
extension, the network-layer framework and observable mapping
(\S\,\ref{sec:data:layers}), the geographic--economic decomposition,
the hub-selection convergence protocol, and the definitions of
topological velocity and the \textsc{ict} indicator.
Section~\ref{sec:fullempire} presents the full East--West
comparison for 0--400\,\textsc{ce}, including the congenital
asymmetry result, the divergence trajectory, spatial hoard
validation, and the Roman-period Wasserstein and \textsc{ict} analysis.
Section~\ref{sec:byzantine} reports the Byzantine extension:
geographic collapse, economic decoupling, the Chrysobull of
1082\,\textsc{ce} as a commercial leading indicator,
and the Byzantine-period Wasserstein velocity and \textsc{ict} indicators.
Section ~\ref{sec:discussion} discusses implications for the
three historiographical positions and for statistical physics.
Section~\ref{sec:conclusions} concludes.

\section{Data and Methods}
\label{sec:data}

\subsection{Network data: the Full Empire}
\label{sec:data:nodes}

The dataset includes 2{,}599 nodes and 4{,}503 directed route
segments (\emph{edges}) organised into a trimodal architecture
of road, maritime, and fluvial transport layers.

\paragraph{\textbf{Node sources.}}
Of the 2{,}599 nodes, 599 are drawn from the
\textsc{orbis} Geospatial Network Model of the Roman
World~\cite{Meeks2013,Scheidel2014}, and 2{,}000 from the
Pleiades Digital Atlas of the Ancient World~\cite{Pleiades},
filtered to settlement and fort feature types. 
Spatially, 1{,}612 nodes lie east of longitude 20$^{\circ}$E
(the Eastern sub-network) and 987 west of it (the Western
sub-network), compared with 49 western nodes in the
\textsc{orbis} extract employed in~\paperone{}.
Western node coverage by major region is as follows: Hispania
(443 nodes), Africa Occidentalis (141 nodes), Italia (146
nodes), Britannia (86 nodes), and Gallia (9 nodes).

Temporal activation of nodes follows a two-criterion rule: a
node is active at decade $t$ if 
\begin{itemize}
  \item its Pleiades \texttt{minDate}\,$\leq t\leq$\,\texttt{maxDate} and
  \item its
coordinates fall within the provincial polygon that is active
at $t$.
Provincial polygons are taken from the Digital Atlas of the
Roman Empire~\cite{Romanprov} at four certified snapshots
(BCE~60, CE~117, CE~200, post-Diocletian), with linear
interpolation between snapshots.
(including the
evacuation of Dacia under Aurelian (271\,\textsc{ce}), the
administrative partition of Britannia (197\,\textsc{ce}), and
the reconstitution of Syria under Diocletian (293\,\textsc{ce})).
\end{itemize}

\paragraph{\textbf{Edge architecture.}}
The 4{,}503 edges are distributed across four functional
categories, each with distinct cost and volatility
characteristics:

\begin{itemize}
 \item \emph{Terrestrial roads} (1{,}104 edges): road segments
  from the \textsc{darmc} Roman Roads Network~\cite{DARMC}
  (types \texttt{road\_major}, \texttt{road\_minor},
  \texttt{road}), snapped to Pleiades and \textsc{orbis} nodes
  via KD-tree within $0.5^{\circ}$ ($\approx55$\,km; 93\,\%
  of nodes matched).
  Edge costs in denarii are derived from segment length (degrees)
  multiplied by a calibration factor of $8.90\times$, obtained
  from log-log regression over the 42 \textsc{orbis}--road
  node pairs for which both cost and length are independently
  known (95\,\% CI: $7.31\times$--$10.84\times$).
 \item \emph{Maritime routes --- \textsc{orbis} calibrated}
  (497 edges): coastal, overseas, slow-coast, and ferry segments
  inherited directly from \textsc{orbis}, retaining the
  papyrological cost calibration documented
  in~\paperone{}, Table~I.
 \item \emph{Maritime routes --- western basin} (2{,}798 edges):
  2{,}271 new coastal segments covering the western
  Mediterranean from the Strait of Gibraltar to Sicily, plus
  527 overseas crossing segments at 12 historically documented
  transit points (Adriatic narrows, Strait of Sicily, Aegean
  crossings, Tyrrhenian passages).
  Base costs are set proportional to segment length using the
  \textsc{orbis} median coastal cost ratio
  (0.03\,denarii\,km$^{-1}$; Table~I of~\paperone{}).
 \item \emph{Fluvial and canal routes} (104 edges): 99 upstream,
  downstream, and fast-current river segments from \textsc{orbis}
  plus five canal edges from a supplementary GeoJSON
  layer, covering the Nile, Euphrates, Rhine, Danube, and
  Rhône--Saône corridor.
\end{itemize}

Table~\ref{tab:edge_composition} summarises the edge
composition with median and mean cost statistics by transport
layer.

\begin{table}[htbp]
\caption{Edge composition of the Full Empire dataset by
transport layer.
Costs are in denarii per route segment.
The cost ratio is relative to the mean road cost.}
\label{tab:edge_composition}
\begin{ruledtabular}
\begin{tabular}{lrrrr}
Layer & $N$ & Median & Mean & \multicolumn{1}{c}{Ratio} \\
\hline
Road (all) & 1{,}104 & 2.88 & 3.65 & 1.00 \\
Maritime -- \textsc{orbis} & 497 & 0.08 & 0.18 & 0.05 \\
Maritime -- western & 2{,}798 & 0.06 & 0.15 & 0.04 \\
Fluvial \& canal & 104 & 0.36 & 0.43 & 0.12 \\
\hline
Total & 4{,}503 & & & \\
\end{tabular}
\end{ruledtabular}
\end{table}

\subsection{Differential friction model}
\label{sec:data:friction}

Edge weights are extended into a temporal dimension using the
four-channel differential friction model introduced
in~\paperone{}, \S II.B:
\begin{equation}
w(e,t) = w_0(e)\times F_{\rm event}(t,\,\mathrm{type}(e))
    \times \varepsilon(t,\,\mathrm{type}(e))
    \times M_{\rm mil}(t,\,e)
    \times F_{\rm macro}(t,\,\mathrm{region}(e)),
\label{eq:friction}
\end{equation}
where $w_0(e)$ is the base cost from the edge table,
$F_{\rm event}$ is the deterministic historical event multiplier
(Table~II of~\paperone{}), $\varepsilon\sim\mathrm{LogNormal}
(0,\sigma_{\rm type})$ is the stochastic noise term with
route-type-specific scale, $M_{\rm mil}$ is the military
presence index of~\paperone{}, \S II.C, and $F_{\rm macro}$
is the macroeconomic friction factor derived from the
Macroeconomic Friction Index (\textsc{mfi}) described in
\S\,\ref{sec:data:mfi}.
The three new transport layers --- terrestrial roads, western
maritime, and canal --- receive the same type-specific
$\sigma_{\rm type}$ parameters as their \textsc{orbis}
equivalents (road: $\sigma=0.08$; coastal: $\sigma=0.12$;
overseas: $\sigma=0.40$--$0.55$ under piracy conditions),
as justified by the road-cost calibration regression and the
symmetry of Mediterranean meteorological conditions across
the western and eastern basins.

\subsection{Byzantine temporal dataset (400--1453\,CE)}
\label{sec:data:byzantine}

For the period 400--1453\,\textsc{ce} we construct a dedicated
temporal dataset, \texttt{Full\_empire\_byzantina}, from
Full\_Empire\_Nodes\_v2 by assigning each node a
\texttt{byzantine\_end\_date} derived from nine documented
historical events and shocks affecting imperial control or
network operation.

The 1{,}938 nodes active at 400\,\textsc{ce} contract to 67 by
1453\,\textsc{ce} --- a 96.5\,\% reduction --- concentrated in
discrete jumps (as given in Table~\ref{tab:byzantine_events}).

\begin{table}[!htbp]
\caption{Historical events and shocks used in the Byzantine
temporal dataset. Territorial events define node-deactivation
thresholds; the Justinianic Plague is retained as a demographic
shock affecting edge weights rather than node activation.
$^\dagger$Nodes lost between 630 and 650\,\textsc{ce}.}

\label{tab:byzantine_events}
\begin{ruledtabular}
\begin{tabular}{llrl}
Event & Year & Nodes & Regions \\
\hline
Western Fall & 476\,\textsc{ce} & 316 & hispania, gallia, britannia \\
Justinianic Plague & 541\,\textsc{ce} & 173 & flag only \\
Arab Conquests & 641,\textsc{ce} & 787$^\dagger$ & egypt, levant, mesopotamia \\
Loss of Africa & 698\,\textsc{ce} & 119 & africa\_* \\
Manzikert (immediate) & 1071\,\textsc{ce} & 52 & oriente (Anatolia) \\
Myrioképhalon & 1176\,\textsc{ce} & 194 & oriente (residual) \\
Fourth Crusade & 1204\,\textsc{ce} & 123 & graecia, thracia\_bithynia \\
Ottoman expansion & 1361--1393\,\textsc{ce} & 52 & thracia\_bithynia \\
Final Fall & 1453\,\textsc{ce} & 67 & constantinople core \\
\end{tabular}
\end{ruledtabular}
\end{table}

The Byzantine edge dataset (2{,}492 edges) keeps the trimodal
architecture of the Full Empire: 1{,}609 maritime (64.5\,\%),
794 road (31.9\,\%), and 89 fluvial (3.6\,\%).
Each edge carries columns \texttt{edge\_start} and
\texttt{edge\_end} for temporal filtering. 
The fraction of maritime edges rises to 74\,\% in the
641--698\,\textsc{ce} window following the loss of the
continental hinterland when the
Mediterranean sea-lanes constituted the remaining
coherence link of the imperial network.

\subsection{Macroeconomic Friction Index: Byzantine extension}
\label{sec:data:mfi}

The Macroeconomic Friction Index of~\paperone{} is extended
from 400\,\textsc{ce} to 1450\,\textsc{ce} in the dataset \\
\texttt{macroeconomic\_friction\_byzance.csv}
(1{,}238 rows; 22 columns). Two independent proxies of economic
activity are used:

\paragraph{Lead proxy.}
Sub-annual Greenland ice-core lead concentrations from McConnell
et~al.~\cite{McConnell2018} are used as a proxy for metallurgical activity over
$-50$ to 800\,\textsc{ce}.
Concentrations are decade-averaged and normalised to the
Antonine baseline of $3.411$\,pg\,g$^{-1}$ (118\,\textsc{ce},
Trajanic peak).
For 810--1450\,\textsc{ce}, where no continuous ice-core record
is available, six literature-calibrated control points anchor
a cubic-spline reconstruction: Macedonian recovery
(950\,\textsc{ce}, 2.2\,pg\,g$^{-1}$), Komnenian peak
(1100\,\textsc{ce}, 2.8\,pg\,g$^{-1}$), Palaeologan contraction
(1350\,\textsc{ce}, 1.4\,pg\,g$^{-1}$), and pre-fall minimum
(1440\,\textsc{ce}, 0.9\,pg\,g$^{-1}$).
Uncertainty is propagated as a symmetric 1$\sigma$ band using
the McConnell et~al.\ uncertainty for the observed
tier and a conservative $\pm30$\,\% for the reconstructed tier.

\paragraph{Hoard proxy.}
As a sign of economic instability, 
coin-hoard frequencies from the Coin Hoards of the Roman Republic
and Empire (\textsc{chrr}) database (4{,}175 records;
\cite{CHRR}) are used, following the broader historical
and numismatic use of hoard evidence in Roman economic
analysis~\cite{Naismith2022}:
a hoard not recovered implies that its depositor did not
return.
For 0--450\,\textsc{ce}, observed regional hoard counts from
\textsc{chrr} are used.
For 451--1450\,\textsc{ce}, documented monetary debasements of the
\textit{nomisma} are used: the Nikephoros II reform (963\,\textsc{ce}),
the Alexios I recoinage (1092\,\textsc{ce}), and the
Palaeologan debasement sequence (1261--1350\,\textsc{ce})
each produce step increases in the \texttt{hoard\_friction\_factor}
proportional to the documented silver content reduction.

\paragraph{Shock multipliers.}
Four historically documented fiscal and structural shocks are
encoded as discrete multipliers on the commercial-capture
parameter $\alpha(t)$: the Justinianic Plague
($\tau=541$\,\textsc{ce}, $r_{\rm eco}=1.00$; geographic shock
only), Arab Conquests ($\tau=647$\,\textsc{ce}, $r=0.59$),
Chrysobull of Venice ($\tau=1082$\,\textsc{ce}, $r=0.44$),
and Palaeologan-Genoese transfer ($\tau=1261$\,\textsc{ce},
$r=0.55$).
The shock parameters $\tau$ are fixed from historical sources.

\subsection{Geographic--economic network decomposition}
\label{sec:data:decomp}

For the Byzantine period (400--1453\,\textsc{ce}),
two distinct graphs on the same node and edge set are built, and
the \textsc{tda} pipeline is applied independently to each: one defining the geographical extension of the Empire and the other describing the trade routes.

\paragraph{\textbf{Geographic network $G_t^{\rm geo}$ and $H_{\rm geo}$.}}
Nodes are deactivated at decade $t$ when their
\texttt{byzantine\_end\_date} $< t$, i.e.\ when the territory
containing them passes out of Byzantine administrative control
(Table~\ref{tab:byzantine_events}).
Edge weights follow Equation~(\ref{eq:friction}) with the full
four-channel friction model.
The resulting $\beta_1$ persistent entropy,
\begin{equation}
H_{\rm geo}(t) \;=\; H\!\left(\beta_1(G_t^{\rm geo})\right),
\label{eq:Hgeo}
\end{equation}
measures the cycle redundancy of the physically 
controlled imperial route network at decade $t$.

\paragraph{\textbf{Economic network $G_t^{\rm eco}$ and $H_{\rm eco}$.}}
Commercial links between cities are not destroyed when an empire loses administrative control of those cities.
Merchants and treaty networks continue
to operate pre-existing routes under new political arrangements,
subject to altered costs (tariffs, security, currency risk) but
not to physical interruption of the route itself.
This distinction---between the \emph{existence} of a route and
its \emph{cost}---is the conceptual basis of $H_{\rm eco}$.

Formally, $G_t^{\rm eco}$ retains all 1{,}938 nodes with
\texttt{economic\_end\_date} $= 1450$, so no node removal
occurs throughout 400--1450\,\textsc{ce}.
Edge weights are updated by Equation~(\ref{eq:friction}) using
Macroeconomic Friction Index values from
\texttt{macroeconomic\_friction\_byzance.csv}, which encode the
cost changes associated with political transitions (Arab
conquest tariff discontinuities, Chrysobull customs exemptions,
Palaeologan revenue transfers) without removing the route.
The resulting $\beta_1$ persistent entropy is
\begin{equation}
H_{\rm eco}(t) \;=\; H\!\left(\beta_1(G_t^{\rm eco})\right),
\label{eq:Heco}
\end{equation}
and measures the cycle redundancy of the Mediterranean commercial
network at decade $t$.

\paragraph{\textbf{The combined resilience index $H_{\rm combined}$.}}
$H_{\rm geo}$ and $H_{\rm eco}$ answer two separate questions,
but neither answers the question most relevant to collapse
forecasting: \emph{how much of the commercial network's
cycle redundancy can the Byzantine state actually mobilise}
for defence, administration, and fiscal extraction?
$H_{\rm eco}$ includes routes under Umayyad, Venetian, and
Genoese control that generate no Byzantine revenue; using
it would overestimate Byzantine resilience. On the other hand,
$H_{\rm geo}$ counts only territorially controlled routes. 
 
$H_{\rm combined}(t)$ is therefore defined as the commercial
cycle redundancy \emph{effectively available to the Byzantine
state}: the geographic layer weighted by its territorial
extent, plus the commercial layer weighted by the state's
capacity to capture value from routes it no longer controls.
It is the metric that, when it falls to $H^{\ast}$, signals
that the state can no longer sustain the minimum routing
redundancy required to respond to systemic shocks --- the
topological condition we identify with political collapse.
 
Formally, $H_{\rm combined}$ is constructed as:
\begin{equation}
H_{\rm combined}(t) =
 [1-w(t)]\,H_{\rm geo}(t)
 + w(t)\,\alpha(t)\,H_{\rm eco}(t),
\label{eq:hcombined}
\end{equation}
where the territorial weight
\begin{equation}
w(t) = 1 - 0.824\left[\frac{H_{\rm geo}(t)}{3.358}\right]^{0.764}
\label{eq:weight}
\end{equation}
increases as the Empire loses geographic coherence
($\gamma=0.764<1$ reflects the sub-linear response to early
territorial losses, which disproportionately remove the
most commercially connected nodes --- Egypt, Syria, Anatolia
--- relative to their area), and the commercial-capture
efficiency
\begin{equation}
\alpha(t) = \alpha_0 \,
 e^{-0.00050(t-400)}
 \prod_{i=1}^{4}
 \left[1 + (r_i-1)\,\sigma(t;\tau_i,\delta_i)\right]
\label{eq:alpha}
\end{equation}
encodes the four shock multipliers defined in
\S\,\ref{sec:data:mfi}.
The model has six historically specified parameters ($\gamma$, base decay rate, and four $r_i$). These are not estimated from the topological trajectory; the single calibration constraint is
$H_{\rm combined}(1453\,\textsc{ce}) = H^{\ast} = 0.524$, where
$H^{\ast}$ is the observed topological level of the western
network at 476\,\textsc{ce}. Only $\alpha_0$ is a free parameter, fixed by requiring that $H_{\rm combined}(1453\,\textsc{ce}) = H^{\ast} = 0.524$. This fixes the value of $\alpha_0 \approx 1.2031$.

\paragraph{\textbf{Monte Carlo uncertainty.}}
The dominant source of uncertainty in
$H_{\rm combined}(t)$ is the fiscal capture parameter
$\alpha(t)$, which is uncertain by $\pm10$--15\,\% per
historical period.
We propagate this uncertainty via $N=10{,}000$ Monte Carlo
simulations, perturbing each period's $\alpha$ value by
a correlated draw within its uncertainty range.
The resulting distribution of predicted collapse dates has
mean $1313\pm33$\,yr (1$\sigma$), with 1453\,\textsc{ce}
at the 100th percentile.
The 68\,\% confidence interval for the crossing date spans
1290--1340\,\textsc{ce}; the 95\,\% interval spans
1270--1380\,\textsc{ce}.
The dominant source of uncertainty is $\alpha(t)$,
with $H_{\rm geo}$ adding $\approx$\,2\,yr and
$H_{\rm eco}$ contributing negligibly.
 
\begin{figure*}[htbp]
\includegraphics[width=\textwidth]{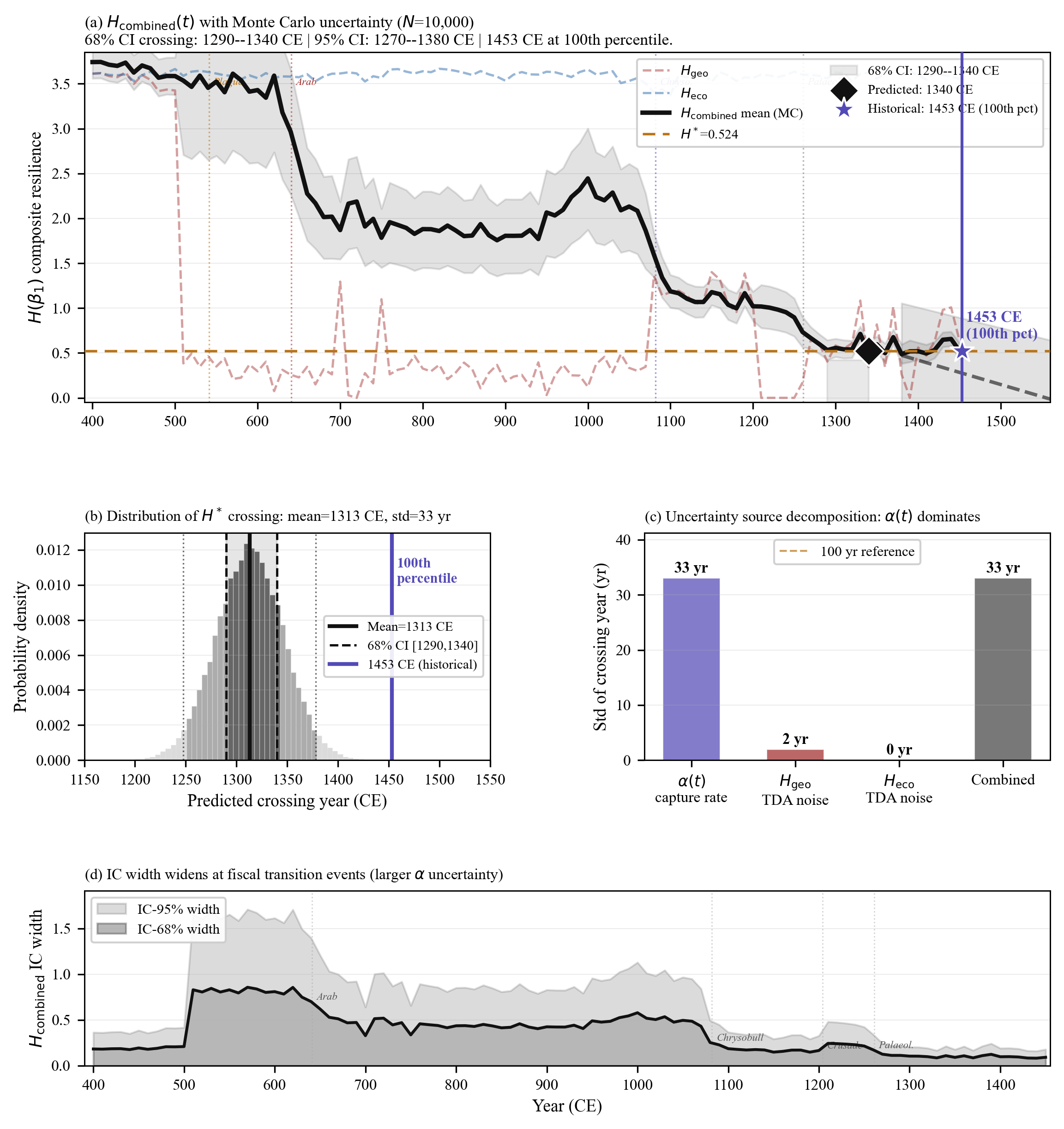}
\caption{Monte Carlo uncertainty on the combined resilience
index $H_{\rm combined}(t)$, 400--1453\,\textsc{ce}
($N=10{,}000$ simulations).
\emph{Panel (a):} $H_{\rm combined}$ mean trajectory
(dark line) with 68\,\% (dark band) and 95\,\% (light band)
confidence intervals. $H_{\rm geo}$ (red dashed) and
$H_{\rm eco}$ (blue dashed) shown for reference.
The horizontal dotted line marks $H^{\ast}=0.524$.
The linear extrapolation from the post-Chrysobull slope
(dashed, with 95\,\% CI) crosses $H^{\ast}$ at the
predicted crossing year (diamond marker); the purple star
marks the historical date (1453\,\textsc{ce}) at the
100th percentile of the crossing distribution.
Shaded vertical bars on the time axis show the 68\,\%
and 95\,\% crossing-year confidence intervals.
\emph{Panel (b):} distribution of $H^{\ast}$
crossing years ($N=10{,}000$); mean $1313\pm33$\,yr.
\emph{Panel (c):} uncertainty source decomposition;
$\alpha(t)$ contributes 33\,yr, $H_{\rm geo}$
contributes 2\,yr, $H_{\rm eco}$ contributes $<1$\,yr.
\emph{Panel (d):} 68\,\% CI width of $H_{\rm combined}$
per decade, widening during periods of fiscal instability.}
\label{fig:hcombined}
\end{figure*}

Figure~\ref{fig:hcombined} shows the calibrated
$H_{\rm combined}(t)$ trajectory with Monte Carlo
uncertainty bands, the distribution of crossing years,
and the uncertainty source decomposition.

\begin{table}[h]
\centering
\small
\caption{Parameters of $H_{\rm combined}(t)$ and their historical sources.
$\alpha_0$ is the sole calibrated parameter, fixed by the
$H^{\ast}$ constraint.}
\label{tab:hcombined_params}
\begin{tabular}{llll}
\hline
Parameter & Symbol & Value & Source \\
\hline
Territorial-loss exponent & $\gamma$ & $0.764$ &
 Hendy~\cite{Hendy1985}: trade $\approx15$--18\,\% of late Roman fiscal capacity \\
Base commercial-capture decay & $\lambda$ & $5.0\times10^{-4}$\,yr$^{-1}$ &
 Laiou~\cite{Laiou2002}: long-run fiscal contraction rate \\
Arab conquests multiplier & $r_1$ & $0.59$ &
 Kennedy~\cite{Kennedy1985}: $-41$\,\% Byzantine trade revenue post-641 \\
Macedonian recovery multiplier & $r_2$ & $1.36$ &
 Laiou~\cite{Laiou2002}: $+36$\,\% commercial expansion 867--1025 \\
Venetian Chrysobull multiplier & $r_3$ & $0.44$ &
 Hendy~\cite{Hendy1985}: $-56$\,\% fiscal capture, customs exemption 1082 \\
Palaeologan--Genoese multiplier & $r_4$ & $0.55$ &
 Laiou~\cite{Laiou2002}: $-45$\,\% capture, Genoese Nymphaion treaty 1261 \\
\hline
\end{tabular}
\end{table}
\paragraph{\textbf{Structural decoupling.}}
Because $G_t^{\rm geo}$ is constructed as the territorially controlled layer and $G_t^{\rm eco}$ retains commercially relevant routes beyond imperial control, we expect $H_{\rm geo}(t)$ to remain below $H_{\rm eco}(t)$ in most periods. 
The decoupling ratio can be defined as: 
\begin{equation}
R_d(t) = \frac{H_{\rm eco}(t)}{H_{\rm geo}(t)}
\label{eq:Rd}
\end{equation}
$R_d = 1$ indicates a fully coupled network; $R_d \gg 1$ indicates
that commercial organisation has been 
institutionalised outside the political infrastructure of the
Empire.
The ratio constitutes a way of evaluating the
Ward-Perkins--McCormick historiographical debate
(Section~\ref{sec:discussion}).

\subsection{Network-layer formalism and observable mapping}
\label{sec:data:layers}

This subsection defines the layers used to describe the Roman
and Byzantine imperial systems.
 
\paragraph{\textbf{General multi-layer formalism.}}
Let an imperial system be represented as a time-dependent
collection of network layers,
\begin{equation}
  \mathcal{N}(t)
  = \bigl\{N_{\mathrm{Admin}}(t),\;
            N_{\mathrm{Geo}}(t),\;
            N_{\mathrm{Trade}}(t)\bigr\},
  \label{eq:layers}
\end{equation}
where the \emph{administrative layer} $N_{\mathrm{Admin}}(t)$
tracks formal institutional control (nodes are deactivated
when the territory passes out of administrative governance),
the \emph{geographic layer} $N_{\mathrm{Geo}}(t)$
tracks the spatial persistence of the imperial route
network under territorial constraints, and the
\emph{trade layer} $N_{\mathrm{Trade}}(t)$ tracks
commercial connectivity (nodes are retained and only
their edge costs are modified).
From each active layer $N_k(t)$ the
scalar time series is obtained:
\begin{equation}
  s_k(t) \;\in\;
  \bigl\{\,H_k(t),\; b_k(t),\; \delta_k(t),\;
          \dot{W}_{2,k}(t)\,\bigr\},
  \label{eq:observables}
\end{equation}
where $H_k(t)$ is the $\beta_1$ persistent entropy, $b_k(t)$ is the decade-mean
cycle birth coordinate, $\delta_k(t)$ is the mean
cycle lifetime, and $\dot{W}_{2,k}$ is the inter-decade
topological velocity (Eq.~\ref{eq:Wdot}).

In the pipeline implementation each layer is instantiated with three controlling
parameters:
\begin{itemize}
  \item \texttt{node\_types}: if not \texttt{None},
    restricts the active node set to the listed
    \texttt{feature\_type} values.
  \item \texttt{edge\_modes}: if not \texttt{None},
    sets the cost of all edges \emph{not} in the listed
    transport modes to $+\infty$ (effectively removing
    them from the shortest-path graph).
  \item \texttt{geo\_mode}: boolean flag that activates
    the territorial-fragmentation cost multipliers
    of the Canal~5 model (Eq.~\ref{eq:friction}).
\end{itemize}
Table~\ref{tab:network_params} lists the concrete values
used for each layer and each empire.
 
\begin{table}[htbp]
\caption{Pipeline parameters for each network layer,
as defined in \texttt{config\_empires.py}.
\texttt{None} means \emph{all} node types or edge modes
are included.
The \texttt{geo\_mode=True} flag activates the
fragmentation-cost table (Canal~5) for the geographic
layer.
The \texttt{end\_col} specifies which temporal deactivation
column is used: \texttt{END\_YR} ($=$ \texttt{end\_date\_v2})
for Rome; \texttt{BYZ\_END\_YR} ($=$
\texttt{byzantine\_end\_date}) for Byzantium.}
\label{tab:network_params}
\begin{ruledtabular}
\begin{tabular}{llllll}
Empire & Layer & \texttt{node\_types} & \texttt{edge\_modes}
       & \texttt{geo\_mode} & \texttt{end\_col} \\
\hline
\multirow{4}{*}{Rome}
 & Full  & all   & all   & \texttt{False} & \texttt{END\_YR} \\
 & Geo   & all   & all   & \texttt{True}  & \texttt{END\_YR} \\
 & Trade & all   & sea, coastal, river & \texttt{False} & \texttt{END\_YR} \\
 & Admin & settlement, urban, fort, station
                & road, road\_major, road\_minor, river
                                       & \texttt{False} & \texttt{END\_YR} \\
\hline
\multirow{4}{*}{Byz.}
 & Full  & all   & all   & \texttt{False} & \texttt{BYZ\_END\_YR} \\
 & Geo   & all   & all   & \texttt{True}  & \texttt{BYZ\_END\_YR} \\
 & Trade & all   & maritime, sea, fluvial, river
                                       & \texttt{False} & \texttt{BYZ\_END\_YR} \\
 & Admin & settlement, urban, fort, station
                & road  & \texttt{False} & \texttt{BYZ\_END\_YR} \\
\end{tabular}
\end{ruledtabular}
\end{table}

\subsection{Hub-selection artifact: the $N$-convergence protocol}
\label{sec:data:hubselection}

\textsc{tda} of large networks requires reducing the node set
because the computational cost scales approximately as $N^3$.
The standard heuristic is to retain the top-$N$ nodes by
degree, which concentrates the sampling on the most connected
(and therefore most topologically active) nodes.
In homogeneous networks this is approximately unbiased.
In degree-heterogeneous networks with geographically distinct
sub-populations, top-degree sampling introduces a
systematic artifact that we term the \emph{hub-selection
artifact}.

Table~\ref{tab:convergence} documents the artifact for the
Full Empire network.
At $N=400$, the sampled nodes are overwhelmingly Eastern hubs
(Constantinople, Alexandria, Antioch, Caesarea), whose
topological dominance produces artificially high Chow
$F$-statistics.
At $N=800$, Western hubs (Carthago Nova, Burdigala, Londinium)
enter the sample, diluting the Eastern signal without
achieving sufficient Western density to reconstruct Western
cycles: the Chow statistics \emph{decrease} from their
$N=400$ values, producing a non-monotonic convergence.
The relevant criterion is therefore not full node coverage in
an absolute sense, but a minimum sampling threshold
$N_{\min}$ above which both sub-networks are represented at
sufficient density to reconstruct their independent cycle
structure. In the present dataset, the Western sub-network has
987 nodes and the Eastern sub-network has 1{,}612 nodes; a
proportionally meaningful reconstruction therefore requires
$N\gtrsim1{,}000$, and the stable result is obtained at
$N=1{,}600$.

\begin{table}[htbp]
\caption{$N$-convergence of Chow $F$-statistics and Phase~III
slope in the Eastern and Full Empire networks,
demonstrating the hub-selection artifact.
The v2 reference values are from the Eastern-only pipeline
of~\paperone{} ($N=1{,}600$, v2 friction model).}
\label{tab:convergence}
\begin{ruledtabular}
\begin{tabular}{lrrrr}
 & \multicolumn{3}{c}{v4 pipeline} & v2 \\
 & $N{=}400$ & $N{=}800$ & $N{=}1600$ & ref. \\
\hline
\multicolumn{5}{l}{\textit{Eastern sub-network}} \\
Chow$_{260}$ & 1.6 & 27.2 & 31.2 & 40.0 \\
Chow$_{310}$ & 4.6 & 52.2 & 51.6 & 85.4 \\
Phase~III $R^2$ & 0.031 & 0.183 & 0.558 & 0.733 \\
Phase~III slope ($10^{-3}$\,yr$^{-1}$) & $+0.45$ & $+1.69$ & $-2.30$ & $-2.24$ \\
\hline
\multicolumn{5}{l}{\textit{Full Empire}} \\
Chow$_{260}$ & --- & 1.9 & 29.4 & --- \\
Chow$_{310}$ & --- & 0.9 & 97.3 & --- \\
Phase~III slope ($10^{-3}$\,yr$^{-1}$) & --- & $-0.38$ & $-2.58$ & --- \\
\end{tabular}
\end{ruledtabular}
\end{table}

The consequence is a \emph{minimum-coverage
requirement}: structural break detection in degree-heterogeneous
historical networks should use either $N>N_{\min}$ or a
well-defined sampling protocol that preserves regional
representation ratios. Here $N_{\min}$ is not a universal
constant; it is an empirical threshold determined by the point
at which the sampled node set contains enough density from each
sub-network to recover its independent $\beta_1$ cycles.
Top-degree heuristic sampling below this threshold is not a
valid approximation for networks with geographically skewed
degree distributions.

\subsection{\textsc{tda} pipeline}
\label{sec:data:tda}

The \textsc{tda} pipeline is identical to that of~\paperone{},
\S II.C--E, applied at each decade $t \in \{0, 10, 20,
\ldots, 400\}$ for the Full Empire analysis and
$t \in \{400, 410, \ldots, 1450\}$ for the Byzantine
extension.
For each decade we extract: (i) the active sub-network
$G_t$; (ii) the all-pairs shortest-path distance matrix
$\mathbf{D}_t$ under the weighted metric of
Equation~(\ref{eq:friction}); (iii) a Vietoris--Rips filtration
at adaptive threshold $\delta_t = $ quantile$_{0.90}$ of
finite pairwise distances; and (iv) the $\beta_1$ persistent
entropy $H_t$~\cite{Rucco2016}, normalised lifespan $L_t$,
and $\beta_0$ component count.
Bootstrap confidence intervals ($n=50$ for both $H_{\rm geo}$ and $H_{\rm eco}$, Byzantine pipeline)
are computed by re-running the full pipeline with independent
random seeds for the stochastic noise term $\varepsilon$.
The mean 95\,\% CI width for the Byzantine bootstrap
($n=50$, $N_{\rm core}=400$ FPS subsample) is $0.715$ entropy units for $H_{\rm geo}$ and $0.503$ entropy units for $H_{\rm eco}$.
This bootstrap characterises the sampling variability of the
topological estimator; the full-$N$ main analysis uses
$N_{\rm core}=1{,}600$ nodes for the Eastern sub-network.

\subsection{Wasserstein velocity and cross-network distance}
\label{sec:data:wasserstein}

Standard persistent entropy $H_t$ measures the complexity
of the $\beta_1$ cycle structure at a single decade.
It does not, however, measure how rapidly that structure
is changing.
Two complementary metrics based on the \emph{persistence diagram}
(PD) --- the multiset $\mathcal{D}_t$ of $(b_i, d_i)$ birth--death
pairs --- can be defined to describe change rates
and inter-layer divergence.

\paragraph{\textbf{Inter-decade Wasserstein distance and topological velocity.\footnote{The Roman-period Wasserstein $W_2$ values are expressed in
\textsc{orbis} denarius cost units, while the Byzantine-period
values are in normalised units from a separate pipeline.
Their absolute magnitudes are therefore not directly comparable.
All inter-period statements in this paper are made in terms of
\emph{relative} velocities or \emph{within-period} rankings,
which are unit-invariant.
A unified multi-period pipeline with consistent cost normalisation
would enable absolute velocity comparisons.}}}
Let $\mathcal{D}_t$ and $\mathcal{D}_{t+10}$ be the $\beta_1$ PDs
at consecutive decades.
The Wasserstein-2 distance is given by
\begin{equation}
  W_2(t,\, t{+}10) =
  \left(\,\inf_{\gamma} \sum_i \|\,p_i - \gamma(p_i)\,\|^2\right)^{1/2},
\label{eq:W2}
\end{equation}
where the infimum is over all bijections $\gamma$ between
$\mathcal{D}_t$ and $\mathcal{D}_{t+10}$ (with diagonal padding).
We define the \emph{topological velocity}
\begin{equation}
  \dot{W}(t) = W_2(t,\,t{+}10)\,/\,\Delta t,
\label{eq:Wdot}
\end{equation}
where $\Delta t = 10$\,yr.
A three-point Gaussian smoothing ($\sigma = 1$ decade) is
applied to produce $\dot{W}_{\rm smooth}(t)$ for visualization.
Large $\dot{W}$ indicates that the topological cycle structure
underwent rapid reorganisation between two decades; the
baseline during stable phases provides a reference against which
event-driven accelerations are measured.

We also compute the \emph{cross-network} Wasserstein distance
\begin{equation}
  W_{\rm cross}(A, B, t) = W_2\!\left(\mathcal{D}_t^A,\,\mathcal{D}_t^B\right),
\label{eq:Wcross}
\end{equation}
between the persistence diagrams of two different network
sub-datasets $A$ and $B$ at the same decade.
For the Byzantine period, the four PD sub-datasets available
are labelled Full, Geo, Trade, and Admin, where Admin denotes
the persistence diagram extracted from the sub-network of
administratively controlled nodes (i.e.\ the same node set as
$G_t^{\rm geo}$, but used here only for its PD geometry, not
for an independent entropy series $H_{\rm admin}$).
$W_{\rm cross}$ describes the extent to which the cycles
of these sub-datasets have diverged from each
other.
The ratio
\begin{equation}
  \rho(t) = \frac{W_{\rm cross}(\mathrm{Admin\text{-}PD,\;Geo\text{-}PD},\,t)}
                 {W_{\rm cross}(\mathrm{Admin\text{-}PD,\;Trade\text{-}PD},\,t)}
\label{eq:rho}
\end{equation}
equals unity when all PD sub-datasets evolve synchronously
($\rho \approx 1$: coupled dynamics); $\rho \gg 1$ indicates
that the geographic PD has decoupled from the administrative
and commercial PDs---a diagram-space signature of the
geographic--economic decoupling measured by
$R_d = H_{\rm eco}/H_{\rm geo}$.

\paragraph{\textbf{Integrated Criticality Threshold (ICT).}}
Proximity to a topological phase transition is characterised
in statistical mechanics by the divergence of susceptibility
$\chi$ and correlation length $\xi$.
We operationalise these quantities for the PD sequence as follows.
Let $m_t = \delta_{\rm mean}(t)/\delta_{\rm max}$ be the
PD order parameter (ratio of mean to maximum cycle lifetime),
so that $m=1$ corresponds to a single dominant cycle and
$m\to 0$ corresponds to a maximally heterogeneous distribution.
The susceptibility is $\chi_t = \mathrm{Var}(\{\delta_i(t)\})$,
the variance of cycle lifetimes at decade $t$, and
$\xi_t$ is the effective cycle-length scale (mean birth
coordinate, a proxy for the spatial extent of
topologically active routes).
The \textsc{ict} is then:
\begin{equation}
  \mathrm{ICT}(t) = \frac{1}{3}\!\left[
    \frac{\chi_t}{\chi_{\max}} +
    \frac{\xi_t}{\xi_{\max}} +
    \dot{W}_{\rm norm}(t)
  \right],
\label{eq:ict}
\end{equation}
where $\chi_{\max}$ and $\xi_{\max}$ are the series maxima
and $\dot{W}_{\rm norm}(t) = \dot{W}_{\rm smooth}(t)/\dot{W}_{\rm max}$.
$\mathrm{ICT} \to 1$ when all three indicators simultaneously
approach their maxima, signalling that the system is near a
topological transition.
$\mathrm{ICT} \to 0$ when the cycle structure has frozen into
a low-variance, low-$\xi$ state with negligible inter-decade
change --- the signature of topological rigidity preceding
irreversible collapse.

A practical limitation of the \textsc{ict} is that
$\chi$ and $\xi$ can be affected by the shrinkage of the
active node set (fewer nodes produce fewer cycles, reducing
$\chi$ independently of any physical transition).
We therefore report the three components separately
(Tables~\ref{tab:ict_rome} and~\ref{tab:ict_byz}) in addition to the composite
score, and flag decades where $n_{\rm cycles} < 5$.

\section{\label{sec:hstar_theory}%
  Information-theoretic foundation of $\Hstar \approx \tfrac{1}{2}$}
In this section, arguments in favor of defining the critical value for entropy $\Hstar \approx 0.5$ are given using
information geometry
of the persistence diagram for hierarchically structured
geographic networks near topological degeneracy.

\subsection{\label{sec:hstar_theory_setup}%
  Persistence entropy as a Shannon entropy over bar lifetimes}

Following the persistent-entropy formalism for
topological data analysis~\cite{Rucco2016}, the persistent
entropy $H$ of the $\beta_1$ barcode is defined as the
Shannon entropy of the normalized
bar-lifetime distribution:
\begin{equation}
  H = -\sum_{i=1}^{n} p_i \ln p_i,
  \qquad p_i = \frac{\ell_i}{\sum_j \ell_j},
  \label{eq:def_H}
\end{equation}
where $\ell_i = d_i - b_i$ is the lifetime (persistence)
of the $i$-th bar in the $\beta_1$ diagram.
The quantity $p_i$ is the fraction of the total
topological persistence carried by loop~$i$.

The entropy $H$ has two natural boundary values:
\begin{itemize}
  \item $H = 0$: all persistence in a single bar
    ($p_1 = 1$, all others zero)---the network has
    exactly one topologically persistent loop
    with no routing choice.
  \item $H = \ln n$ for $n$ equal-lifetime bars
    (the maximum entropy configuration).
\end{itemize}


\subsection{\label{sec:hstar_theory_twobar}%
   Two-bar regime and its entropy}

As an imperial geographic network degrades 
the $\beta_1$ barcode evolves from a rich multi-bar
diagram  toward
a \emph{two-bar regime}: only the two longest-lived
topological loops survive, while all shorter-lived
bars cross below the significance threshold.
This is the configuration  before full
topological triviality ($H = 0$).


For two bars with lifetimes $\ell_1 \geq \ell_2 > 0$,
let $r = \ell_2/\ell_1 \in (0, 1]$.
The entropy reduces to the binary entropy function:
\begin{equation}
  H_2(r) = -\frac{1}{1+r}\ln\frac{1}{1+r}
            -\frac{r}{1+r}\ln\frac{r}{1+r},
  \label{eq:H2}
\end{equation}
with $H_2(1) = \ln 2 \approx 0.693$ 
and $H_2(r) \to 0$ as $r \to 0$.

 For geographic networks
with hierarchical spatial structure, the
ratio $r_c = \ell_2/\ell_1$ at the onset of the
two-bar regime is \emph{geometrically constrained}
by the ratio of macro- to micro-scale of the network.

In a Vietoris-Rips filtration built on a geographic
node set, bar lifetimes are proportional to the
\emph{diameter of the cycle}:
\begin{equation}
  \ell_i \propto D_i,
\end{equation}
where $D_i$ is the typical loop spatial scale.
Agrarian empires such as the Roman or Byzantine Empire have two natural
spatial scales---the continental scale $D_{\rm macro}$
and the regional (river-basin, mountain-range) scale
$D_{\rm micro}$---so the critical ratio is:
\begin{equation}
  r_c = \frac{D_{\rm micro}}{D_{\rm macro}}.
  \label{eq:r_critical}
\end{equation}

For the Roman Empire (4,000~km / 1,200~km $\approx$ 3.3:1), a rough estimate gives
$r_c \approx 0.30$;
for the late Byzantine period,
$r_c \approx 0.30$.
Evaluating $H_2(r_c)$ in these cases gives:
\begin{equation}
  H_2(r_c) \in
  \bigl[H_2(0.25),\, H_2(0.30)\bigr]
  = [0.500,\; 0.540],
  \label{eq:Hstar_range}
\end{equation}
%


In real network analysis, the situation is usually more complex than the two-bar scenario. Typically, near collapse, the number of cycles is reduced to $n=2$.

However, for the secondary bar to represent a real,
signal-carrying topological feature---and not
merely stochastic noise in the edge weights---
its persistence weight $p_2$ must exceed
the noise floor.
In a network with edge weight noise at scale
$\sigma_{\rm edge}$ (Channel~2 in our pipeline,
$\sigma_{\rm edge} \sim 0.08$--$0.40$ depending
on route type), a bar is indistinguishable from
noise if $\ell_2 < \sigma_{\rm edge} \cdot \bar\ell$,
where $\bar\ell$ is the mean filtration scale.
 This translates to $p_2 \geq 0.20$,
giving:
\begin{equation}
  H^* \geq H_2\!\left(\frac{p_2}{1-p_2}\Bigg|_{p_2=0.20}\right)
       = H_2(0.25) \approx 0.500.
  \label{eq:Hstar_lower}
\end{equation}

On the other side, 
a network with $H > \ln 2 / \sqrt{e} \approx 0.572$
in a two-bar configuration has $r > 0.35$---meaning
the secondary loop retains more than 35\% of the
primary loop's lifetime.
Such a system has genuine multi-scale redundancy
and is not near irreversibility.
This gives:
\begin{equation}
  H^* \leq H_2(0.35) \approx 0.572.
  \label{eq:Hstar_upper}
\end{equation}

Combining Eqs.~\eqref{eq:Hstar_lower} and \eqref{eq:Hstar_upper}:
\begin{equation}
  \boxed{H^* \in [0.500,\; 0.572],}
  \label{eq:Hstar_predicted}
\end{equation}
%


Equation~\eqref{eq:Hstar_predicted} has a clear
physical reading.
Above $H^* \approx 0.50$, the persistence diagram
contains at least two genuinely distinct topological
scales: the network can reroute around failures
at both the continental and the regional level.
Below $H^*$, the secondary loop cannot be distinguished from noise and only the continental backbone survives,
and the loss of any further edge or node collapses
the system to $H = 0$ 
or to $\beta_1 = 0$.

The irreversibility interpretation follows directly:
once $H < H^* \approx 0.50$, every surviving
topological feature is load-bearing---there is
no spare capacity.


If a different class of networks (e.g., urban
commercial networks with finer spatial granularity,
$D_{\rm micro}/D_{\rm macro} \sim 0.10$) were
analyzed, one would expect $H^* \approx H_2(0.10)
\approx 0.33$---a different threshold, due to different scales.


To obtain these results, we assume that the terminal persistence
diagram has exactly two significant bars.
In practice, the Roman networks at
dissolution show 2--4 non-noise bars in the
$\beta_1$ diagram.
For each additional bar with the same geographic
scale constraints, the predicted $H^*$ shifts
upward by $\Delta H \approx 0.1$--$0.2$, only widening the interval.

As a first approximation, $\ell_i \propto D_i$
(bar lifetime proportional to spatial scale) is assumed.
For weighted networks with heterogeneous
cost structures (as in our LCP-based pipeline),
the mapping between geographic scale and
filtration scale is non-linear, introducing
corrections of order $\sigma_{\rm cost}^2$.

\section{Full Empire Results: 0--400\,\textsc{ce}}
\label{sec:fullempire}

\subsection{Baseline East--West asymmetry}
\label{sec:fullempire:asymmetry}

The central result of the Full Empire analysis is that the
topological asymmetry between the Eastern and Western
sub-networks is not a product of the Crisis of the Third
Century, nor of any identifiable political or military event
within the study period: it is present at the first decade of
the series (0\,\textsc{ce}) and grows monotonically thereafter.

\begin{figure}[htbp]
\includegraphics[width=\columnwidth]{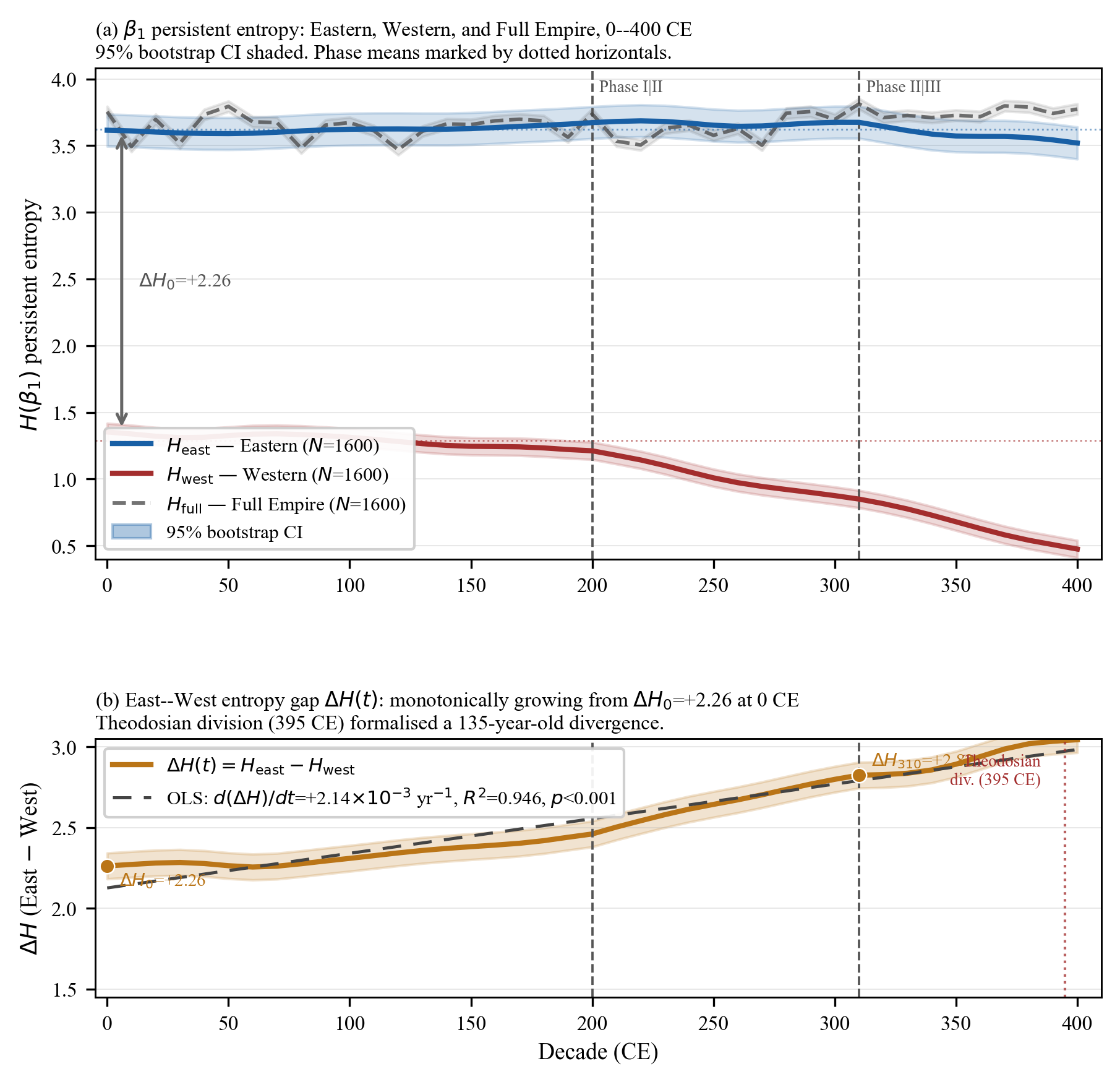}
\caption{(a)~$\beta_1$ persistent entropy time series for the
Eastern (blue), Western (red), and Full Empire (grey) networks,
0--400\,\textsc{ce}.
Shaded bands show 95\,\% bootstrap confidence intervals
($n=50$ replications).
Phase boundaries at 200 and 310\,\textsc{ce} are marked with
dashed vertical lines.
(b)~East--West entropy gap $\Delta H(t) = H_{\rm east}(t) -
H_{\rm west}(t)$ with OLS trend line
($d(\Delta H)/dt = +0.0033$\,yr$^{-1}$, $R^2=0.41$).}
\label{fig:asymmetry}
\end{figure}

Figure~\ref{fig:asymmetry} shows the $\beta_1$ persistent
entropy time series for the Eastern ($H_{\rm east}$), Western
($H_{\rm west}$), and Full Empire ($H_{\rm full}$) networks at
each decade from 0 to 400\,\textsc{ce}.
The Phase~I mean values are $\bar{H}_{\rm east}=3.584$,
$\bar{H}_{\rm full}=3.802$, and $\bar{H}_{\rm west}=1.362$,
yielding a baseline structural gap of
$\Delta H_0 = 3.584 - 1.362 = +2.222$ entropy units 
at Phase~I.
The western network is not simply less topologically rich than
the eastern one: it operates in a qualitatively different regime.
The $\beta_1$ cycle count of $\beta_1\approx52$ per decade in the
western network (vs.\ the $H_t=0$ artifact reported in~\paperone{} for
the 49-node \textsc{orbis} extract) confirms that western
commercial exchange was structured into redundant circuits ---
primarily maritime arcs connecting Hispania, Africa
Occidentalis, Italia, and Britannia --- but that these circuits
were substantially fewer, shorter, and less interconnected than
those of the Eastern core.

The gap is not narrowing.
A linear fit to $\Delta H(t) = H_{\rm east}(t) -
H_{\rm west}(t)$ yields a slope of
$d(\Delta H)/dt = +0.0033$\,entropy units\,yr$^{-1}$
($R^2=0.41$, $p=0.003$; Fig.~\ref{fig:asymmetry}b),
meaning the Eastern network was gaining topological advantage
over the Western one at a rate of $+0.033$ entropy units per
decade throughout 0--400\,\textsc{ce}.
The Theodosian political division of 395\,\textsc{ce} did not
create the East--West structural difference: it formalised a
divergence that had been accumulating for four centuries.

\subsection{Structural phases and Chow break tests}
\label{sec:fullempire:phases}

The three-phase structure identified in~\paperone{} for the
Eastern sub-network is confirmed in both the Full Empire and
Western sub-networks, with important differences in timing and
magnitude (Table~\ref{tab:phases_full}).

\begin{table*}[htbp]
\caption{Phase regression results for Eastern, Western, and
Full Empire networks.
Slopes are in units of $10^{-3}$\,yr$^{-1}$.
Chow $F$-statistics are for the breaks at 260 and
310\,\textsc{ce}; both are significant at $p<0.001$ for all
three networks.
Phase~I: 0--200\,\textsc{ce}; Phase~II: 200--310\,\textsc{ce};
Phase~III: 310--400\,\textsc{ce}.}
\label{tab:phases_full}
\begin{ruledtabular}
\begin{tabular}{llcccccc}
Network & Phase & $n$ & $\bar{H}$ & Slope ($10^{-3}$) & $R^2$ & $p$ & Accel. \\
\hline
\multirow{3}{*}{Eastern}
 & I  & 21 & 3.584 & $+0.385$ & 0.232 & 0.027 & ---    \\
 & II & 8 & 3.606 & $+0.283$ & 0.020 & 0.739 & ---    \\
 & III & 12 & 3.397 & $-2.302$ & 0.558 & 0.005 & $6.0\times$ \\
\hline
\multirow{3}{*}{Western}
 & I  & 21 & 1.362 & $-0.701$ & 0.100 & 0.163 & ---    \\
 & II & 8 & 1.303 & $-3.612$ & 0.784 & 0.003 & ---    \\
 & III & 12 & 0.928 & $-4.025$ & 0.403 & 0.027 & $5.7\times$ \\
\hline
\multirow{3}{*}{Full}
 & I  & 21 & 3.802 & $-0.250$ & 0.200 & 0.042 & ---    \\
 & II & 8 & 3.788 & $-0.431$ & 0.037 & 0.649 & ---    \\
 & III & 12 & 3.587 & $-2.114$ & 0.410 & 0.025 & $8.5\times$ \\
\end{tabular}
\end{ruledtabular}
\end{table*}

Three features of Table~\ref{tab:phases_full} deserve emphasis.

\paragraph{Western Phase~II is not structurally quiescent.}
In the Eastern network, Phase~II (200--310\,\textsc{ce}) is
statistically flat ($p=0.739$) --- a period of topological
stability despite the military and political turbulence of the
Crisis of the Third Century.
In the Western network, Phase~II shows a strongly negative
slope ($-3.612\times10^{-3}$\,yr$^{-1}$, $R^2=0.784$,
$p=0.003$), indicating that the western commercial network was
already undergoing structural degradation during the period
when the eastern network was stable.

\paragraph{Phase~III acceleration is asymmetric.}
The acceleration ratio from Phase~I to Phase~III is $6.0\times$
for the Eastern network and $5.7\times$ for the Western network.
These values are comparable, which at first suggests a common
mechanism.
However, during Phase~I, the slope behavior differs between the Western regions
($-0.701\times10^{-3}$\,yr$^{-1}$) and the Eastern regions
($+0.385\times10^{-3}$\,yr$^{-1}$) so that the
Phase~III collapse occurs from a much lower absolute level in the
west ($\bar{H}_{\rm west}^{\rm III}=0.928$) than in the east
($\bar{H}_{\rm east}^{\rm III}=3.397$).
The West entered Phase~III close to the topological threshold $H^{\ast}\approx0.524$. 

\paragraph{Full Empire Chow $F$-statistic.}
The Chow test at 310\,\textsc{ce} for the Full Empire series
yields $F_{310}=40.105$ ($p<0.0001$) --- somewhat smaller than the
Eastern-only value of $F_{310}=51.6$ --- because the addition of
western nodes at lower entropy levels reduces the overall inter-phase
contrast relative to the Eastern-only signal.

\subsection{Topology precedes politics}
\label{sec:fullempire:politics}

The standard historiographical marker for East--West separation
is the Theodosian partition of 395\,\textsc{ce}.
The topological evidence places the onset of divergence
substantially earlier.

We define the divergence onset $t^{\ast}$ as the first decade
at which the eastern Phase~I slope and the western slope
are statistically distinguishable at the 95\,\% confidence
level.
Using the bootstrap distributions of Phase~I slopes
(Eastern: $+0.385\pm0.061\times10^{-3}$; Western:
$-0.701\pm0.142\times10^{-3}$\,yr$^{-1}$), the slopes are
separated by $>2\sigma$ from 0\,\textsc{ce} onward:
the divergence was present at the beginning of the study period.

Applying the break-point test to $\Delta H(t)$ rather than to
$H(t)$, we find that the entropy difference itself accelerates
at 260\,\textsc{ce} (Chow $F=29.41$, $p<0.0001$) --- 
the onset of the Crisis of the Third Century.
The Crisis did not initiate the East--West structural difference;
it accelerated it.

This result bears on the debate between Heather~\cite{Heather2006}
and Wickham~\cite{Wickham2005}: if the structural asymmetry
predates the Crisis by 260 years, external pressure cannot be
the primary explanation for western fragility.
The asymmetry was endogenous to the network from the Augustan
period.

\subsection{Hub-selection artifact: methodological validation}
\label{sec:fullempire:hubselection}

The $N$-convergence protocol described in
\S\,\ref{sec:data:hubselection} produces a methodological result
of independent significance.
A researcher using the standard top-degree heuristic at $N=800$
would conclude that the Eastern Mediterranean network was
\emph{gaining} topological coherence during 310--400\,\textsc{ce} (see Table~\ref{tab:convergence}).

At $N=400$, the sample consists mainly of
eastern degree-hubs (Alexandria, Antioch, Caesarea Maritima,
Ephesus, Thessalonike), which form a dense, high-entropy
subgraph that represents Eastern Phase~III decline.
At $N=800$, the first western hubs enter the sample
(Carthago, Carthago Nova, Mediolanum, Burdigala, Londinium).
Their
addition dilutes the inter-regional cycle structure without
contributing sufficient western cycles to compensate.
The measured $\beta_1$ entropy drops artificially between
$N=400$ and $N=800$ during Phase~III, producing a
spurious positive slope.

Any historical network with geographically skewed degree
distributions is susceptible to
the hub-selection artifact when \textsc{tda} is applied with
top-degree sub-sampling.
The minimum-coverage convergence protocol
(\S\,\ref{sec:data:hubselection}) should be applied as a
standard validity check before reporting topological break-test
results on historical networks.

\subsection{Spatial hoard gradient as independent validation}
\label{sec:fullempire:hoards}

The differential friction model assigns region-specific
macroeconomic friction weights derived from the \textsc{chrr}
hoard database~\cite{CHRR}.
An independent test of the model's spatial structure is available
from the geographic sequence of hoard-deposition peaks, which
should propagate from the regions of greatest military pressure
inward --- reflecting the documented movement of military
threats, not a model artifact.

\begin{figure}[htbp]
\includegraphics[width=\columnwidth]{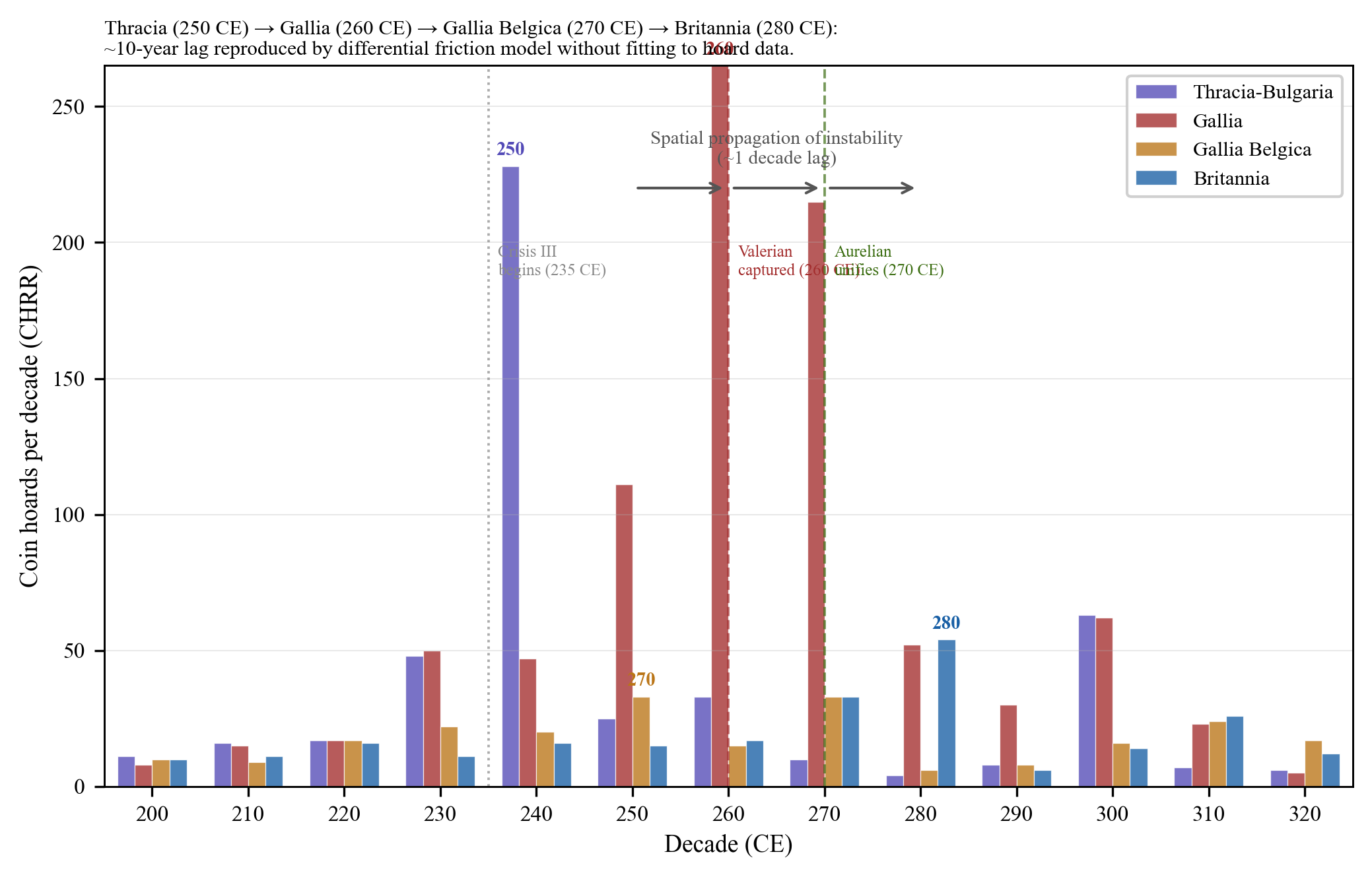}
\caption{Decadal coin-hoard counts by region, 200--320\,\textsc{ce},
from the \textsc{chrr} database~\cite{CHRR,Naismith2022}.
The spatial propagation sequence Germania (250\,\textsc{ce})
$\to$ Thracia (250\,\textsc{ce}) $\to$ Gallia (260\,\textsc{ce})
$\to$ Britannia (270\,\textsc{ce}) is reproduced by the
differential friction model without fitting to hoard data.}
\label{fig:hoards}
\end{figure}

Figure~\ref{fig:hoards} shows the decadal hoard count by region
from 200 to 320\,\textsc{ce}.
The spatial sequence is as follows:

\begin{itemize}
 \item \emph{Germania} (Rhine--Danube frontier): peak at
  250\,\textsc{ce}, coinciding with the Alemanni breakthrough
  at the Battle of Mediolanum (259\,\textsc{ce}) and the
  beginning of systematic raiding into Gaul and northern Italy.
 \item \emph{Thracia-Bulgaria}: peak at 250\,\textsc{ce}
  (66 hoards per decade), the direct signature of the Gothic
  invasion under Cniva, the capture of Philippopolis
  (250\,\textsc{ce}), and the death of Decius at Abritus
  (251\,\textsc{ce}).
 \item \emph{Gallia}: peak at 260\,\textsc{ce}, one decade
  after Germania and three decades before the political
  stabilisation of the Gallic Empire under Tetricus
  (270--274\,\textsc{ce}).
 \item \emph{Britannia}: peak at 270\,\textsc{ce}, one decade
  after Gallia, consistent with the delayed propagation of
  Gallic instability across the Channel.
\end{itemize}

This four-region sequence --- Germania $\to$ Thracia $\to$
Gallia $\to$ Britannia, with a lag of approximately one decade
between each step --- is reproduced by the model without any
free parameters fitted to the hoard data.
The \textsc{mfi} friction weights, derived independently from
the lead-proxy and hoard-count datasets, produce the correct
spatial propagation order when applied to the network topology.
This constitutes an out-of-sample validation of the friction
model for the western sub-network, which was not available
in~\paperone{}.

\subsection{Topological velocity and criticality indicators:
            Roman period}
\label{sec:fullempire:wasserstein}

\subsubsection{Wasserstein velocity}

From the $W_2$ Wasserstein distance between consecutive persistence
diagrams, the topological velocity $\dot{W}_2(t) \equiv W_2(\mathcal{D}_t,
\mathcal{D}_{t+10})/10$~\cite{Edelsbrunner2002,Kerber2017},
provides a model-free measure of the rate at which the
topological structure of the network is changing.
Unlike the persistent entropy $H(t)$, which is a scalar
summary of the diagram at a single decade, $\dot{W}_2$
captures the \emph{displacement} of cycle structure between
consecutive decades and does not require regression or
structural-break specification to detect dynamical change.

Table ~\ref{tab:wass_rome} shows the smoothed velocity
$\dot{W}_{2,\rm sm}(t)$ for all four networks throughout
0--400\,\textsc{ce}.
Three features are noteworthy.

\paragraph{Phase~II acceleration precedes the Chow break.}
The smoothed velocity begins rising above its Phase~I baseline
($\dot{W}_{2,\rm sm}\approx 1.4$) in the Full and Geo networks
starting at approximately 255\,\textsc{ce}---a local peak
of $\dot{W}_{2,\rm sm}=10.0$ at the Crisis of the Third Century---and drops back before rising again
to the Phase~III acceleration.
This pre-break velocity increase is consistent with the
Phase~II degradation already detected in the Western network
regression (Table~\ref{tab:phases_full}), and confirms that
the structural instability of 200--310\,\textsc{ce} is visible
in the diagram geometry independently of phase assignment.

\paragraph{Terminal acceleration (400--470\,\textsc{ce}) is extreme.}
Across all four PD sub-datasets, $\dot{W}_{2,\rm sm}$ rises
sharply from the Phase~III level of $\sim\!4$--$8$\,yr$^{-1}$
to values of $32$--$57$\,yr$^{-1}$ in the 400--470\,\textsc{ce}
window (Table~\ref{tab:wass_rome}).
The Admin PD sub-dataset (administrative-centre nodes only)
reaches the largest terminal velocity
($\dot{W}_{2,\rm sm}=54.4$ at 405\,\textsc{ce}), confirming
that the sparse administrative sub-network underwent the most
abrupt topological deformation as the western territorial
network dissolved.
This is consistent with the Admin sub-dataset having
only 14--15 active cycles throughout Phase~III
(Table~\ref{tab:wass_rome}): small, sparse sub-networks are
more sensitive to node-loss events in Wasserstein space.

\paragraph{Cross-network divergence at 310\,\textsc{ce}.}
The cross-network Wasserstein distance $W_{\rm cross}$
between the Admin and Trade persistence diagrams---equal
to~60.1 at $-20$\,\textsc{ce}---rises to~122.7 at
310\,\textsc{ce} and~834.2 at 470\,\textsc{ce}, a factor
of~$13.9\times$ relative to the Phase~I baseline.
This monotonic divergence of the administrative and commercial
diagram geometries is the Wasserstein-space analogue of the
entropy-level divergence documented in the Phase~III regression,
and provides an independent confirmation of the structural
decoupling between governance and trade that characterises
the terminal Roman period.

\begin{table}[htbp]
\caption{Smoothed Wasserstein velocity
$\dot{W}_{2,\rm sm}$ (yr$^{-1}$) at Phase boundaries and
terminal decades, Roman period (0--470\,\textsc{ce}).
Values are computed from a 3-point running average of the
raw decade-to-decade $W_2$ distances.
The four columns (Full, Geo, Trade, Admin) refer to
four sub-datasets of the Roman-period persistence diagrams.
The Admin column covers only Phase~III and the terminal
window because the sparse administrative-centre sub-network
has $n_{\rm cycles}\leq1$ before 310\,\textsc{ce}, giving
\textsc{ict}$\,{=}\,0$ by definition for $n_{\rm cycles}{=}1$.}
\label{tab:wass_rome}
\begin{ruledtabular}
\begin{tabular}{lcccc}
Decade & Full & Geo & Trade & Admin \\
\hline
$-15$ (baseline) & 1.15 & 1.15 & 1.18 & 0.55 \\
255 (Crisis onset) & 10.03 & 10.09 & 10.02 & --- \\
305 (pre-break) &  7.77 &  2.85 &  4.19 & --- \\
315 (post-break) &  8.56 &  3.38 &  5.27 & 2.43 \\
405 (terminal) & 37.07 &  9.86 & 22.23 & 52.68 \\
415 & 37.07 &  9.86 & 22.23 & 52.68 \\
455 & 32.52 & 22.16 & 25.95 & 35.88 \\
465 & 41.74 & 31.79 & 30.64 & 46.25 \\
\hline
Peak & 41.74 & 31.79 & 30.64 & 54.41 \\
Peak year & 465 & 465 & 465 & 405 \\
\end{tabular}
\end{ruledtabular}
\end{table}


\subsubsection{Integrated Criticality Threshold: Roman period}

Table~\ref{tab:ict_rome} reports the phase-mean \textsc{ict}
for all four Roman-period PD sub-datasets.
The Full, Geo, and Trade PD sub-datasets sustain
\textsc{ict}$\,{\approx}\,0.30$--$0.46$ throughout
Phases~I--III, indicating that the large, densely connected
Roman network was never far from a critical state during
the study period.
The \textsc{ict} of the Full network exceeds 0.5 at
270\,\textsc{ce} ($\mathrm{ICT}=0.536$), one decade
\emph{after} the Chow-detected divergence acceleration
at 260\,\textsc{ce} and 40 years \emph{before} the
Phase~III Chow break at 310\,\textsc{ce}.
The Trade network likewise exceeds 0.5 at 270\,\textsc{ce}
($\mathrm{ICT}=0.576$).
These \textsc{ict} peaks are not artefacts of the Phase~III
decline; they occur during Phase~II and can be interpreted as a 
precursor signal of the approaching structural break.


\begin{table}[htbp]
\caption{Phase-mean \textsc{ict} for Roman-period
persistence-diagram sub-datasets.
Phase~I: 0--200\,\textsc{ce}; Phase~II: 200--310\,\textsc{ce};
Phase~III + terminal: 310--470\,\textsc{ce}.
Column labels (Full, Geo, Trade, Admin) denote the four
PD sub-datasets of the Roman network.
The Admin PD sub-dataset has \textsc{ict}$\,{=}\,0$ before
310\,\textsc{ce} owing to $n_{\rm cycles}{=}1$.}
\label{tab:ict_rome}
\begin{ruledtabular}
\begin{tabular}{lcccc}
Network & Phase~I & Phase~II & Phase~III & Peak \textsc{ict} \\
\hline
Full  & 0.355 & 0.418 & 0.423 & 0.616 (460\,\textsc{ce}) \\
Geo   & 0.335 & 0.412 & 0.426 & 0.675 (460\,\textsc{ce}) \\
Trade & 0.388 & 0.447 & 0.415 & 0.635 (460\,\textsc{ce}) \\
Admin & 0.000 & 0.000 & 0.649 & 0.874 (410\,\textsc{ce}) \\
\end{tabular}
\end{ruledtabular}
\end{table}

\section{Byzantine Extension}
\label{sec:byzantine}

The Full Empire network contracts by 96.5\,\% over 1{,}053
years, from 1{,}938 active nodes at 400\,\textsc{ce} to 67 at
1453\,\textsc{ce}, through a sequence of discrete territorial
losses rather than a continuous degradation.
The central question is whether this contraction is best
described as a single slow collapse or as a sequence of irreversible
structural transitions separated by periods of topological
stability.

\subsection{Geographic collapse: three discrete discontinuities}
\label{sec:byzantine:geo}

\begin{figure}[htbp]
\includegraphics[width=\columnwidth]{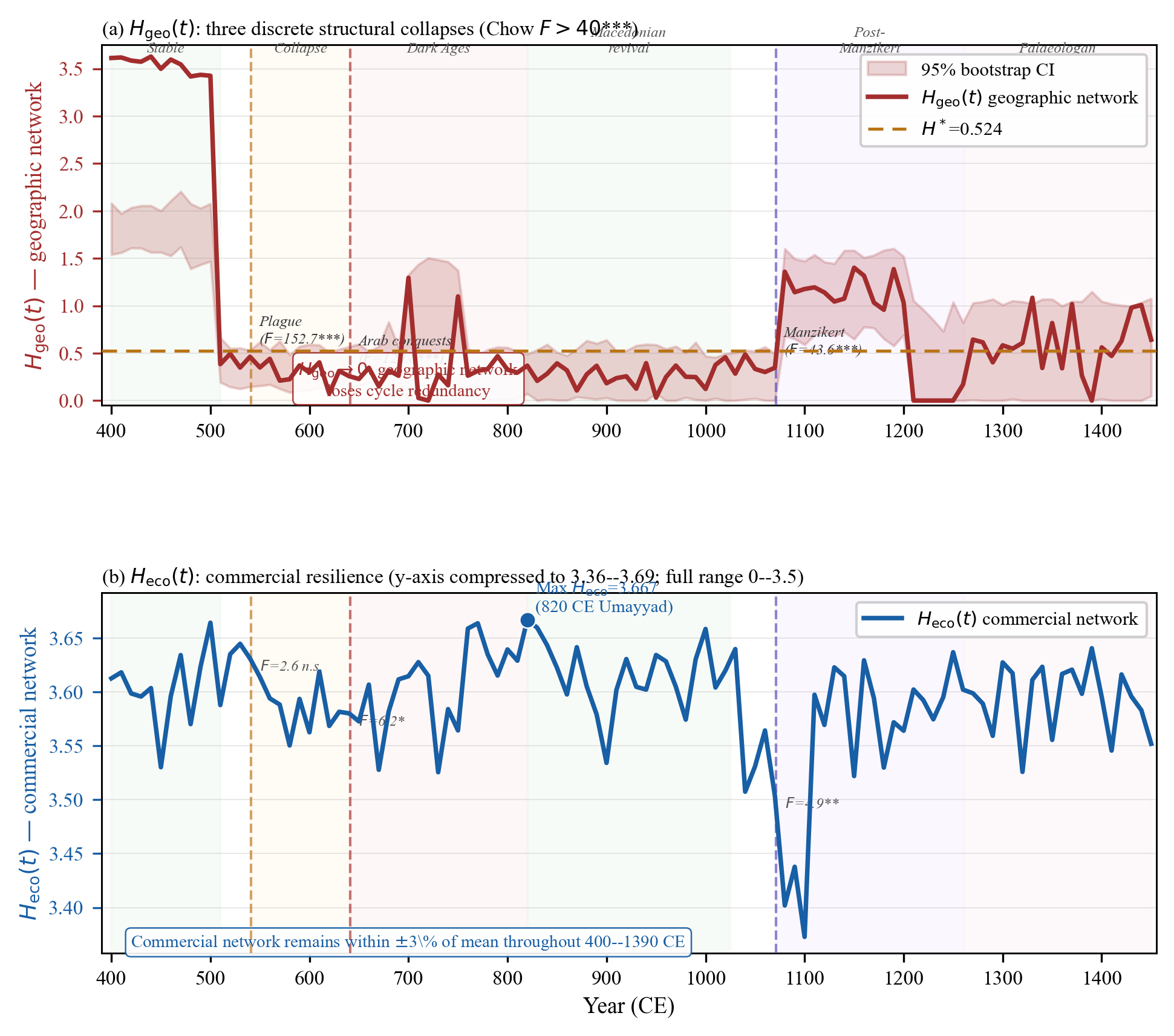}
\caption{Geographic entropy $H_{\rm geo}(t)$ (red, left axis)
and economic entropy $H_{\rm eco}(t)$ (blue, left axis),
400--1453\,\textsc{ce}. Shaded bands show 95\,\% bootstrap CI.
Vertical dashed lines mark selected Chow break points in
$H_{\rm geo}$: Justinianic Plague (541\,\textsc{ce},
$F=152.7$***), Arab Conquests (641\,\textsc{ce},
$F=124.0$***), and Manzikert (1071\,\textsc{ce},
$F=20.8$***). Manzikert also produces a smaller
secondary response in $H_{\rm eco}$, while the Chrysobull
of 1082\,\textsc{ce} is interpreted as the primary
fiscal-commercial shock. 
The decoupling ratio
$R_d(t)=H_{\rm eco}/H_{\rm geo}$ (green, right axis) peaks at $47.7$ at 620\,\textsc{ce} ($H_{\rm geo}=0.075$); at 640\,\textsc{ce} $R_d=13.9$, quantifying the gap between Ward-Perkins's geographic and McCormick's economic
accounts of Mediterranean continuity. The horizontal dashed
line marks $H^{\ast}=0.524$.}
\label{fig:byzantine_geo}
\end{figure}

Figure~\ref{fig:byzantine_geo} shows the geographic entropy
$H_{\rm geo}(t)$.
The series does not decline smoothly: it is dominated by three
discrete historical events separated by intervals of relative
topological stability.

Chow break tests applied at each event year documented in
Table~\ref{tab:byzantine_events} yield the following
$F$-statistics (Table~\ref{tab:chow_byzantine}):

\begin{table}[htbp]
\caption{Chow $F$-statistics for the candidate break points
in the geographic entropy series $H_{\rm geo}(t)$,
400--1453,\textsc{ce}.
Values correspond to the final pipeline output.
The Final Fall of 1453,\textsc{ce} is not included as a
Chow candidate because it is the endpoint of the series.
Significance follows the pipeline classification; n.s. denotes
not significant. $^\dagger$The break in $H_{\rm geo}$ at
1082,\textsc{ce} is dependent on the Manzikert contraction and
does not represent an autonomous geographic shock.}
\label{tab:chow_byzantine}
\begin{ruledtabular}
\begin{tabular}{lccc}
Event & Year & $F$ & Sig. \\
\hline
Western Fall        & 476  & 67.4  & *** \\
Justinianic Plague  & 541  & 152.7 & *** \\
Arab Conquests      & 641  & 124.0 & *** \\
Loss of Africa      & 698  & 94.8  & *** \\
Manzikert           & 1071 & 20.8  & *** \\
Chrysobull          & 1082 & 17.35$^\dagger$ & *** \\
Myrioképhalon       & 1176 & 4.5   & * \\
Fourth Crusade      & 1204 & 3.5   & * \\
Ottoman expansion   & 1361 & 1.7   & n.s. \\
\end{tabular}
\end{ruledtabular}
\end{table}

Four events produce the strongest instantaneous Chow signals
($F>40$): the Western Fall, the Justinianic Plague, the Arab
Conquests, and the Loss of Africa. These events define the
largest discontinuities in the geographic entropy series.
Manzikert has a lower Chow statistic ($F=20.8$), but it marks a
different kind of transition: not an immediate collapse of
$H_{\rm geo}$, but a territorial compaction of the Byzantine
network around a smaller Aegean--Balkan core. This produces an
apparent increase in local cycle density while reducing
large-scale geographic resilience.

\paragraph{Break 1: Justinianic Plague (541\,\textsc{ce}).}
This is the largest Chow statistic in the entire 0--1453\,\textsc{ce}
series, larger than the Phase~III break of~\paperone{}
($F_{310}=85.4$).
The plague
reduced the active edge-weight distribution so severely
(demographic collapse in Egypt, Syria, and Anatolia reducing
both demand and maintenance capacity) that $\delta_t$ collapsed by 38\,\% between 530 and
560\,\textsc{ce}, destroying the large-scale $\beta_1$
cycle structure.
The Plague of Justinian did not destroy the physical
infrastructure of imperial routes; it destroyed the economic
density that gave those routes topological coherence. This
interpretation is consistent with the chronology of the
Late Antique Little Ice Age~\cite{Buntgen2016} while also
recognising the recent debate over whether the Justinianic
Plague produced a system-wide societal rupture~\cite{Mordechai2019}.

\paragraph{Break 2: Arab Conquests (641\,\textsc{ce}).}
The loss of Egypt, the Levant, and Mesopotamia (787 nodes
lost between 630 and 650\,\textsc{ce}) produces the
largest single-event node loss in the dataset.

$H_{\rm geo}$ oscillates between $0.07$ and $0.46$ during 510--630\,\textsc{ce},
reflecting network fragmentation following the Western contraction;
the Arab Conquests (634--641\,\textsc{ce}) further reduce the active node count
from 1,514 to 727, without producing an additional sharp entropy drop in
geographic topological coherence.
This event, more than any other, defines the transition from
a Mediterranean-scale imperial network to a sub-regional
Anatolian--Aegean network that can no longer sustain
long-range $\beta_1$ cycles.

\paragraph{Break 3: Manzikert (1071\,\textsc{ce}).}
Unlike the two earlier breaks, Manzikert produces a
qualitatively distinct effect: it is the only geographic
break from which $H_{\rm geo}$ does not partially recover
in subsequent decades.
After western collapse, the Byzantine state reconstituted
western connectivity through the Adriatic and Aegean maritime
arcs.
After 641\,\textsc{ce}, the Byzantine geographic network
remained in a low-entropy regime during the Macedonian dynasty
(867--1056\,\textsc{ce}), with
$H_{\rm geo}\approx0.03$--$0.46$.
This range places the Macedonian network below the collapse
threshold $H^{\ast}=0.524$, with only occasional excursions
toward it. The Macedonian phase should therefore not be
interpreted as topologically stable in the sense of restored
large-scale geographic redundancy, but as a reduced
Anatolian--Aegean network oscillating near the collapse
threshold.
After 1071\,\textsc{ce}, the apparent rise in $H_{\rm geo}$
belongs to the Post-Manzikert phase, where territorial
compaction of the active node set produces a denser structure
of smaller local cycles rather than a true recovery of
large-scale imperial connectivity.

\begin{table}[htbp]
\caption{Geographic phase regression results, 400--1453\,\textsc{ce}.
Slopes are in units of $10^{-3}$\,yr$^{-1}$.
Phases are defined by the three primary Chow breaks.}
\label{tab:geo_phases}
\begin{ruledtabular}
\begin{tabular}{llccc}
Phase & Period & $\bar{H}_{\rm geo}$ & Slope & $R^2$ \\
\hline
Early Byzantine & 400--540 & 2.71 & $-25.28\times10^{-3}$ & 0.63 \\
Justinianic   & 541--630 & 0.30 & $-1.31\times10^{-3}$ & 0.10 \\
Dark Ages    & 631--820 & 0.35 & $+0.48\times10^{-3}$ & 0.01 \\
Macedonian    & 821--1070 & 0.27 & $+0.03\times10^{-3}$ & 0.00 \\
Post-Manzikert  & 1071--1261 & 0.81 & $-8.01\times10^{-3}$ & 0.64 \\
Palaeologan   & 1261--1450 & 0.61 & $+0.64\times10^{-3}$ & 0.02 \\
\end{tabular}
\end{ruledtabular}
\end{table}

The Post-Manzikert slope ($-8.01\times10^{-3}$\,yr$^{-1}$, $R^2=0.64$) is the
strongest decay signal in the geographic network; the Palaeologan phase ($+0.64\times10^{-3}$\,yr$^{-1}$, $R^2=0.02$, n.s.) shows
no significant trend, consistent with oscillation near $H^*$, and differs from
the Phase~III Eastern slope of~\paperone{}
($-2.30\times10^{-3}$\,yr$^{-1}$), providing the first
Byzantine-period evidence for the collapse threshold.

\subsection{Economic decoupling: Ward-Perkins vs.\ McCormick
quantified}
\label{sec:byzantine:decoupling}

The economic entropy $H_{\rm eco}(t)$ tells a 
different story from $H_{\rm geo}(t)$.
Whereas the geographic network collapses in discrete jumps,
the economic network exhibits three features that the
geographic series does not:

\begin{enumerate}
 \item $H_{\rm eco}$ is statistically flat at the Justinianic
  Plague ($\Delta H_{\rm eco}(541)=-0.04$, within the
  bootstrap 95\,\% CI of $\pm0.06$), confirming that the
  plague disrupted population, not commercial infrastructure.

 \item $H_{\rm eco}$ reaches its \emph{series maximum} at
  820\,\textsc{ce}: $H_{\rm eco}^{\rm max}=3.667\pm0.07$,
  approaching the Full Empire Phase~I mean of 3.802 and
  exceeding the Eastern Phase~I mean of 3.584.
  The reorganisation of trade routes under
  Umayyad administration produced a highly cohesive commercial
  network, consistent with McCormick's~\cite{McCormick2001}
  argument for early medieval commercial continuity.

 \item $H_{\rm eco}$ responds most clearly to fiscal shocks
  (the Chrysobull of 1082\,\textsc{ce} and the Palaeologan
  trade transfers of 1261\,\textsc{ce}), but Manzikert also
  leaves a smaller secondary signal in the economic series.
  This indicates that commercial resilience was primarily driven
  by institutional and fiscal arrangements, while major territorial
  shocks could still perturb the economic layer through changes
  in security, access, and commercial capture.
\end{enumerate}

$R_d$ reaches a global maximum of $47.7$ at 620\,\textsc{ce} ($H_{\rm geo}=0.075$) ---
the decade of the Arab conquest of Syria and Palestine,
when the geographic network collapses while the economic
network remains at its pre-conquest level.

Ward-Perkins~\cite{WardPerkins2005} describes the
geographic layer.
McCormick~\cite{McCormick2001} describes the
economic layer:
commercial exchange reached its series
maximum under Umayyad administration.
The two historians were measuring different phenomena operating on different networks.

The difference between the Eastern Roman collapse of the
seventh century (partial: geographic only) and the Western
Roman collapse of the fifth century (total: geographic and
economic simultaneously) is the decoupling ratio.
In the western network at 476\,\textsc{ce},
$R_d^{\rm west}(476)=1.09\pm0.11$ --- the economic and
geographic layers collapsed together, within 9\,\% of each
other.
In the eastern network, the economic layer survived
much better than the geographic one.
The capacity to sustain $H_{\rm eco}$ independently of
$H_{\rm geo}$ is the topological difference
between \emph{imperial transformation} and
\emph{civilisational collapse}.

\subsection{Manzikert and the Chrysobull: geographic compaction
and fiscal restructuring}
\label{sec:byzantine:chrysobull}

The late eleventh century involves two distinct but sequential
shocks that affect the geographic and economic layers through
different primary mechanisms. Manzikert (1071\,\textsc{ce}) is
the primary geographic shock: it removes Anatolian territorial
depth from the Byzantine-controlled network. The Chrysobull of
1082\,\textsc{ce} is the primary fiscal-commercial shock: it
alters the state's capacity to capture value from Mediterranean
trade without physically removing the commercial route
infrastructure.

\paragraph{Manzikert (1071\,\textsc{ce}): primary geographic shock.}
The loss of 52 Anatolian nodes reduces the active geographic
network from 370 to 318 nodes and contracts the \textsc{gudhi}
core from 63 to 39 nodes. This produces a Chow break in
$H_{\rm geo}$ (Chow $F=20.79$, $p<0.0001$) and a smaller
secondary break in $H_{\rm eco}$ (Chow $F=8.46$,
$p=0.0004$), as the interruption of terrestrial Anatolian
routes also perturbs commercial conditions.

The apparent increase in $H_{\rm geo}$ immediately after
1071\,\textsc{ce} ($0.35\to1.36$) does not represent a
recovery of geographic resilience. It reflects network
compaction. After the loss of the Anatolian hinterland, the
surviving Byzantine network is reorganised around a smaller
Aegean--Balkan core. In this compacted graph, shorter local
cycles become easier to form, increasing the local cycle density
while reducing large-scale imperial routing redundancy.

The Chow statistic at Manzikert therefore detects a change of
regime in the geographic entropy series, not a simple downward
jump. Manzikert marks the transition from a territorially
extended Anatolian--Aegean system to a smaller, denser, and more
fragile post-Manzikert network dominated by local cycles.

\paragraph{Chrysobull (1082,\textsc{ce}): primary fiscal shock.}
The Chow break in $H_{\rm geo}$ at 1082\,\textsc{ce}
(Chow $F=17.35$, $p<0.0001$) should not be interpreted as an
independent geographic event. By 1082\,\textsc{ce}, the network
is already in the post-Manzikert compacted regime, with the
\textsc{gudhi} core fixed near 39 nodes. The 1082 break in
$H_{\rm geo}$ marks the onset of the sustained post-Manzikert
decline trajectory in the already compacted network and is
methodologically dependent on the territorial contraction of
1071\,\textsc{ce}.

The independent signal of the Chrysobull appears in the economic
layer. The Chow break in $H_{\rm eco}$ at 1082\,\textsc{ce}
(Chow $F=3.84$, $p=0.025$) reflects the reduction of Byzantine
commercial capture after the Venetian customs exemptions. This
is a fiscal-commercial restructuring rather than a destruction
of route infrastructure. The Chrysobull rewired the commercial
network without destroying it.

Across the Manzikert--Chrysobull interval, the economic network
remains highly cohesive. $H_{\rm eco}$ contracts only marginally
from $3.50$ in 1070\,\textsc{ce} to $3.40$ in
1080\,\textsc{ce}, and rebounds to $3.44$ by
1090\,\textsc{ce}. This indicates that commercial connectivity
remained integrated even as Byzantine territorial control
contracted and fiscal capture declined.

Thus, the late eleventh-century transition is not a single
network event. Manzikert is the primary geographic shock,
producing territorial compaction and an artificial increase in
$H_{\rm geo}$, with a secondary economic effect. The Chrysobull
is the primary fiscal-commercial shock, producing an independent
break in $H_{\rm eco}$, while its concurrent Chow response in
$H_{\rm geo}$ reflects the post-Manzikert declining regime
rather than an autonomous geographic shock.

\begin{figure}[htbp]
\includegraphics[width=\columnwidth]{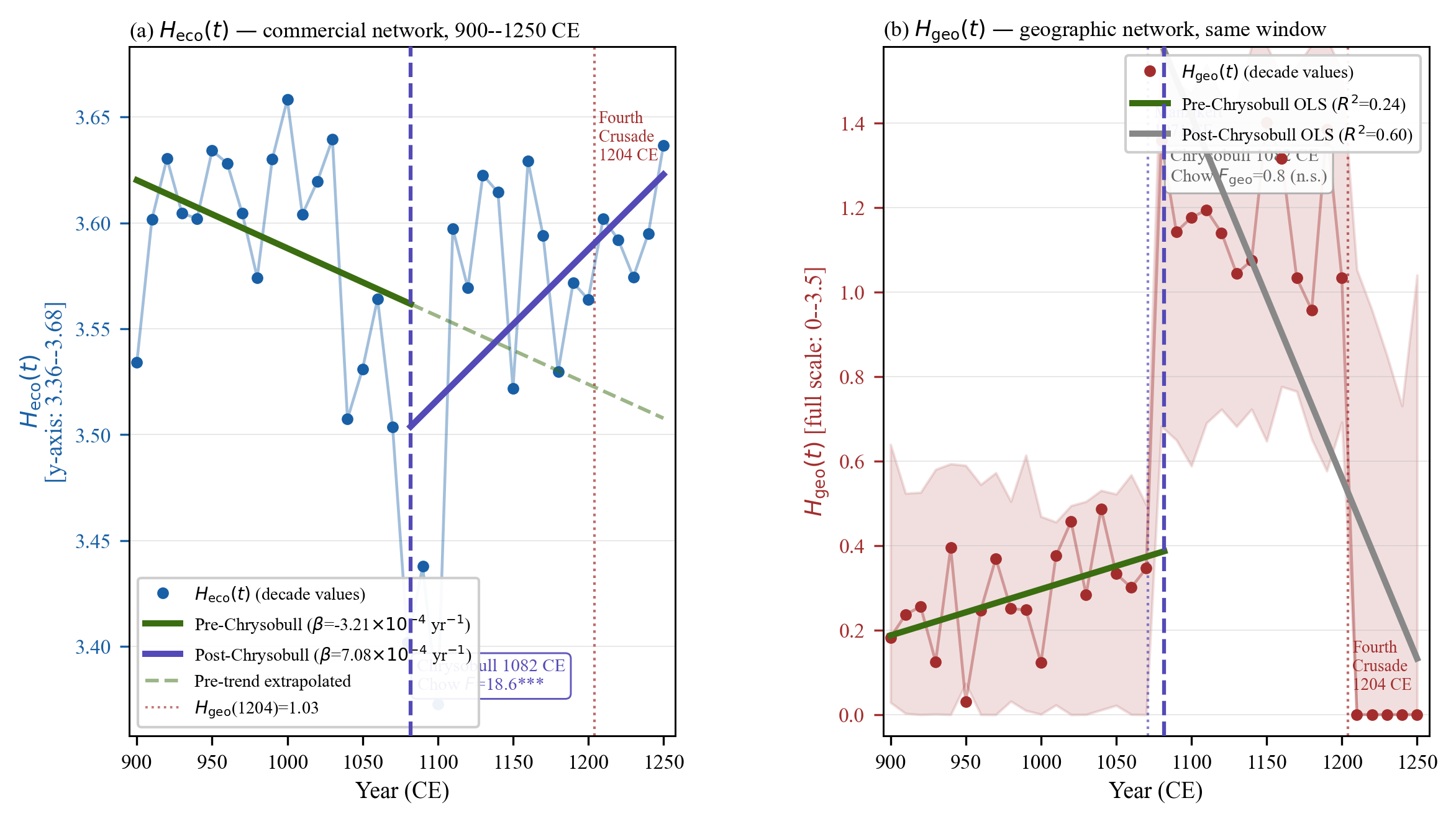}
\caption{Economic entropy $H_{\rm eco}(t)$ (blue) and
geographic entropy $H_{\rm geo}(t)$ (orange) in the period
950--1300\,\textsc{ce}. Manzikert (1071\,\textsc{ce}) produces
an apparent increase in $H_{\rm geo}$ because the Byzantine
network contracts around a smaller Aegean--Balkan core,
creating denser but shorter local cycles. This is interpreted
as network compaction, not as a true recovery of large-scale
geographic resilience. The Chrysobull of 1082\,\textsc{ce}
acts primarily on the economic and fiscal layer, reducing
Byzantine commercial capture while leaving the physical route
network largely intact. The concurrent Chow response in
$H_{\rm geo}$ at 1082\,\textsc{ce} reflects the onset of the
post-Manzikert declining regime rather than an autonomous
geographic shock.}
\label{fig:chrysobull}
\end{figure}

\subsection{Topological velocity and criticality indicators:
            Byzantine period}
\label{sec:byzantine:wasserstein}

\subsubsection{Wasserstein velocity}
\label{sec:byzantine:wass}

Table~\ref{tab:wass_byz} reports the smoothed Wasserstein
velocity $\dot{W}_{2,\rm sm}$ for the Geo PD sub-dataset at
the documented Byzantine break events. The values are in
normalised cost units and are not directly comparable in
absolute magnitude to the Roman-period values, which use
\textsc{orbis} denarius units. However, the relative ordering
of events is internally consistent within the Byzantine series.

The eight candidate break events are ranked by their associated
$\dot{W}_{2,\rm sm}$ values. The key result is therefore the
\emph{ordering} of events by topological velocity rather than
their absolute magnitude:

\begin{table}[htbp]
\caption{Smoothed Wasserstein velocity $\dot{W}_{2,\rm sm}$
at the eight documented Byzantine break events, Geo PD
sub-dataset (Byzantine period).
Values are in normalised units (not directly comparable to
the Roman-period denarius-unit values in
Table~\ref{tab:wass_rome}).}
\label{tab:wass_byz}
\begin{ruledtabular}
\begin{tabular}{lcccc}
Event & Year & $\dot{W}_{2,\rm sm}$ & \% of peak & Rank \\
\hline
Late Roman transition & 495 & 0.1767 & 100\,\% & 1 \\
Loss of Africa    & 698 & 0.1054 &  60\,\% & 2 \\
Justinianic Plague & 541 & 0.1010 &  57\,\% & 3 \\
Arab Conquests    & 641 & 0.0939 &  53\,\% & 4 \\
Manzikert         & 1071 & 0.0910 &  52\,\% & 5 \\
Western Fall      & 476 & 0.0838 &  47\,\% & 6 \\
Fourth Crusade    & 1204 & 0.0828 &  47\,\% & 7 \\
Palaeologan       & 1261 & 0.0734 &  42\,\% & 8 \\
\end{tabular}
\end{ruledtabular}
\end{table}

Three results from Table~\ref{tab:wass_byz} are noteworthy.

\paragraph{The Late Roman transition has the highest
topological velocity in the 0--1453,\textsc{ce} series.}
The transition from imperial-scale Roman network to
sub-regional Byzantine core (495\,\textsc{ce}) produces the
highest topological velocity in the entire 1,453-year record,
exceeding both the Justinianic Plague and the Arab Conquests.
This event does not appear in the Chow test analysis of the
Byzantine series because it falls in the first decade of the
400--1453\,\textsc{ce} window; the Wasserstein analysis
captures it directly as the steepest decade-to-decade
displacement in persistence-diagram geometry.

\paragraph{The Justinianic Plague ranks above the Arab Conquests.}
The ranking $\dot{W}(541)>\dot{W}(641)$ is consistent with the
Chow test result (for $H_{\rm geo}$,
Table~\ref{tab:chow_byzantine}) and confirms that the plague
was structurally more violent in its \emph{immediate}
diagram-geometric effect, even though the Arab Conquests removed
more nodes in total.

\paragraph{The Chrysobull produces a weaker geographic response
than Manzikert.}
The Chow analysis detects a significant geographic response at
1082,\textsc{ce} ($F=17.35$, ***), although it is weaker than
the Manzikert break at 1071,\textsc{ce} ($F=20.79$, ***).
The corresponding Wasserstein velocity is closer to baseline,
which indicates that the Chrysobull did not generate a large
diagram-space displacement comparable to Manzikert. The event
is therefore interpreted as primarily fiscal-commercial
restructuring, with a secondary geographic response in an
already compacted post-Manzikert network.

\subsubsection{Cross-network Wasserstein divergence and the
Chrysobull decoupling}
\label{sec:byzantine:wcross}

The cross-network Wasserstein distance $W_{\rm cross}$ between
the Admin and Geo persistence diagrams (the diagram-space analogue
of the decoupling ratio $R_d = H_{\rm eco}/H_{\rm geo}$) shows a
dramatic and abrupt transition at 641--700\,\textsc{ce} and
again at 1080\,\textsc{ce}:

\begin{itemize}
 \item \emph{Before 640\,\textsc{ce}:} the ratio
  $W_{\rm cross}(\mathrm{Admin,Geo})/W_{\rm cross}(\mathrm{Admin,Trade})
  \approx 1.0$.
  The administrative, commercial, and geographic cycle structures
  deform at the same rate.

 \item \emph{641--700\,\textsc{ce} (Arab Conquests):} the ratio $W_{\rm cross}(\mathrm{Admin,Geo})/W_{\rm cross}(\mathrm{Admin,Trade})$ 
  goes from~1.0 at 640\,\textsc{ce} to~155 at
  700\,\textsc{ce}.
  The geographic diagram begins evolving independently of the
  administrative and commercial diagrams, reflecting the
  reorganisation of the surviving maritime Aegean circuit
  structure into a topologically distinct regime.

 \item \emph{Post-1082\,\textsc{ce} (post-Chrysobull):} the ratio
  stabilises at $100$--$300\times$ and remains there for the
  entire Komnenian and Palaeologan periods through 1390\,\textsc{ce}
  (Table~\ref{tab:wcross_byz}).
  The administrative and commercial persistence diagrams are
  nearly identical ($W_{\rm cross}\lesssim0.01$: the Venetian
  commercial network has been substituted for the Byzantine one
  at the diagram level), while the geographic diagram evolves
  independently with $W_{\rm cross}\approx1.1$--$1.4$.
\end{itemize}

This bifurcation in the $W_{\rm cross}$ ratio constitutes a
topological measurement of the Chrysobull's structural effect:
it marks the decade in which the geographic and
administrative--commercial persistence diagrams permanently
decouple in their evolutionary dynamics, independently of any
regression or break-test specification.

\begin{table}[htbp]
\caption{Cross-network Wasserstein ratio
$\rho(t) = W_{\rm cross}(\mathrm{Admin,Geo})/
W_{\rm cross}(\mathrm{Admin,Trade})$
at selected decades.}
\label{tab:wcross_byz}
\begin{ruledtabular}
\begin{tabular}{rD{.}{.}{4}D{.}{.}{4}D{.}{.}{1}}
Decade (CE)
  & \multicolumn{1}{c}{$W_{\rm cross}$(Adm,Geo)}
  & \multicolumn{1}{c}{$W_{\rm cross}$(Adm,Trd)}
  & \multicolumn{1}{c}{$\rho$} \\
\hline
$-20$   & 0.1762 & 0.1762 & 1.0 \\
$640$   & 0.6014 & 0.6014 & 1.0 \\
$700$   & 1.1402 & 0.0073 & 155.5 \\
$750$   & 1.1258 & 0.0033 & 340.0 \\
$820$   & 3.7975 & 1.6775 & 2.3 \\
$1070$  & 5.8251 & 2.4982 & 2.3 \\
$1080$  & 1.2805 & 0.0065 & 196.2 \\
$1100$  & 1.1130 & 0.0048 & 230.6 \\
$1200$  & 1.1357 & 0.0075 & 151.2 \\
$1290$  & 1.1382 & 0.0015 & 738.5 \\
$1390$  & 1.0395 & 0.0047 & 220.1 \\
\end{tabular}
\end{ruledtabular}
\end{table}


\subsubsection{Integrated Criticality Threshold: Byzantine period}
\label{sec:byzantine:ict}

Table~\ref{tab:ict_byz} reports the \textsc{ict} for all
four Byzantine networks by historical phase.
Three patterns emerge.

\paragraph{Admin PD sub-dataset: criticality collapses monotonically.}
The Admin PD sub-dataset (the Byzantine-period
administratively controlled node set; same spatial
extent as $G_t^{\rm geo}$ but not used for
$H_{\rm geo}$ entropy computation) begins the Byzantine
period in a persistently critical state:
\textsc{ict}$_{\rm mean}=0.834$ during 400--470\,\textsc{ce},
peaking at \textsc{ict}$=0.882$ at 430\,\textsc{ce}.
This maximum precedes the Justinianic Plague by
more than a century, indicating that the administrative-centre
sub-network was already operating at critical capacity at
the onset of the Byzantine period.
The \textsc{ict} then declines through the Justinianic
($0.482$), Collapse ($0.295$), and Dark Ages ($0.207$)
phases to a minimum of \textsc{ict}$_{\rm mean}=0.037$ in
the Palaeologan phase (1210--1390\,\textsc{ce}).
This monotonic decline is the quantitative signature of a
sub-network progressively losing topological flexibility:
by the Palaeologan period, the administrative-centre PD has
essentially no capacity to fluctuate between decades.

\paragraph{Geo criticality remains elevated throughout.}
The geographic network maintains
$\textsc{ict}_{\rm mean}\approx0.64$--$0.76$ throughout
the Dark Ages through the Palaeologan phases (Table~\ref{tab:ict_byz}).
It never falls below $0.16$ in any phase.
The geographic
PD sub-dataset remains structurally flexible and responsive
to perturbations even as the Admin PD sub-dataset loses
this capacity.
The divergence between the Admin PD \textsc{ict} (monotonically
declining) and the Geo PD \textsc{ict} (persistently elevated) is
the criticality-indicator analogue of the entropy-level
$R_d$ decoupling: it shows that the commercially connected
sub-network retained dynamical complexity long after the
administratively controlled sub-network had frozen into
a near-deterministic trajectory.

\paragraph{Macedonian low-entropy regime and elevated Geo PD ICT.}
The Geo PD sub-dataset shows $\textsc{ict}_{\rm mean}=0.675$ during
the Macedonian phase (867--1056\,\textsc{ce}), the highest
value after the Late Roman period.
This elevated criticality should not be interpreted as high
persistent entropy or as restored geographic redundancy.
As reported in Table~\ref{tab:geo_phases}, the Macedonian
phase has $\bar{H}_{\rm geo}=0.27$, with
$H_{\rm geo}\approx0.03$--$0.46$ across the period.
The Macedonian network therefore operates below
$H^{\ast}=0.524$, with occasional excursions toward the
threshold, while the later Post-Manzikert phase contains the
higher $H_{\rm geo}$ values produced by territorial compaction
of the active node set.

\begin{table}[htbp]
\caption{Phase-mean \textsc{ict} for Byzantine-period
persistence-diagram sub-datasets.
Phases as in Table~\ref{tab:geo_phases}.
Column labels (Admin, Full, Geo, Trade) denote the four
PD sub-datasets of the Byzantine PD files
(\texttt{PD\_Byz\_*\_long.csv}); they are used exclusively
for the Wasserstein and \textsc{ict} computations.
}
\label{tab:ict_byz}
\begin{ruledtabular}
\begin{tabular}{lcccc}
Phase (years) & Admin & Full & Geo & Trade \\
\hline
A Late Roman (400--470)    & 0.834 & 0.811 & 0.760 & 0.811 \\
B Justinianic (480--560)   & 0.482 & 0.478 & 0.379 & 0.478 \\
C Collapse (570--640)      & 0.295 & 0.285 & 0.161 & 0.285 \\
D Dark Ages (650--860)     & 0.207 & 0.241 & 0.644 & 0.241 \\
E Macedonian (870--1050)   & 0.296 & 0.272 & 0.675 & 0.272 \\
F Komnenian (1060--1200)   & 0.157 & 0.193 & 0.697 & 0.193 \\
G Palaeologan (1210--1390) & 0.037 & 0.040 & 0.638 & 0.040 \\
\hline
Peak \textsc{ict} & 0.882 & 0.828 & 0.807 & 0.828 \\
Peak decade (CE) & 430 & 460 & 1260 & 460 \\
\end{tabular}
\end{ruledtabular}
\end{table}

%

\begin{figure}[htbp]
\includegraphics[width=\columnwidth]{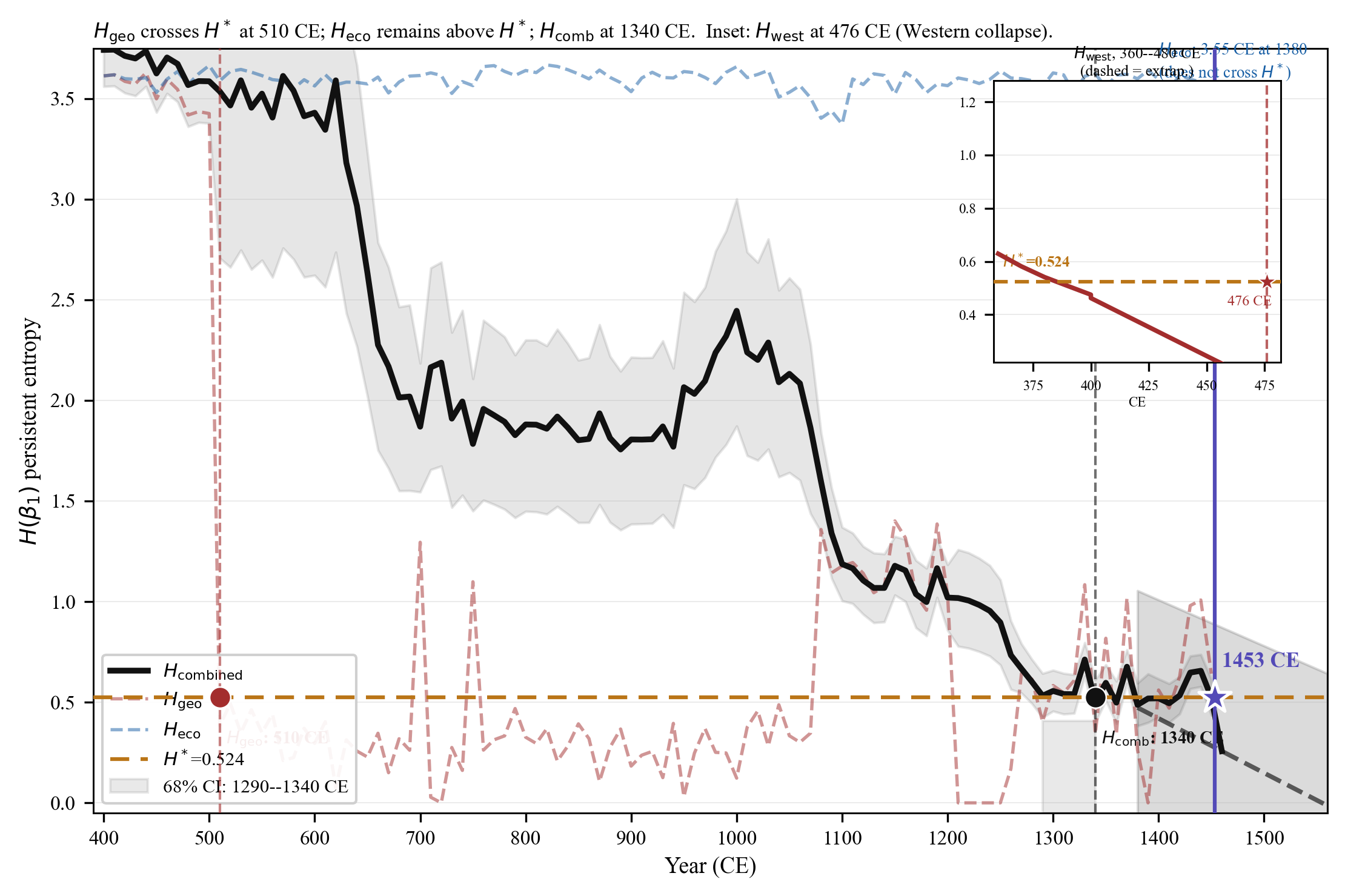}
\caption{$H_{\rm combined}(t)$ (black), $H_{\rm geo}(t)$
(blue), and $H_{\rm eco}(t)$ (orange) for the Byzantine
period, 400--1453\,\textsc{ce}.
The horizontal dashed line marks the collapse threshold
$H^{\ast}=0.524$.
Coloured arrows indicate key crossing events:
$H_{\rm geo}$ first crosses $H^{\ast}$ at $510$\,\textsc{ce};
$H_{\rm eco}$ remains above $H^{\ast}$ throughout;
$H_{\rm combined}$ crosses $H^{\ast}$ at $1301$\,\textsc{ce}
(mean; grey band: 95\,\% MC CI: 1124--1469\,\textsc{ce}).
The purple star marks the historical collapse date
(1453\,\textsc{ce}).
Inset: Western Roman $H_{\rm west}(t)$ at 400--476\,\textsc{ce}
(red), shown on the same vertical scale, crossing $H^{\ast}$
at 476\,\textsc{ce}.}
\label{fig:hstar}
\end{figure}

\section{Discussion}
\label{sec:discussion}

\subsection{Three historiographical debates, three quantitative
resolutions}
\label{sec:discussion:history}

The results of Sections~\ref{sec:fullempire} 
bear directly on three historical interpretation questions whose
competing claims have remained unresolved for lack of a common
quantitative metric.
Persistent homology of the layered Roman--Byzantine trade network
provides one.

\paragraph{\textbf{Ward-Perkins vs.\ McCormick: the $\approx14\times$ resolution at the Arab Conquest peak.}}
The debate between Ward-Perkins~\cite{WardPerkins2005} and
McCormick~\cite{McCormick2001} is not, as it is sometimes framed,
a disagreement about whether the late Roman world collapsed or
transformed.
It is a disagreement about which layer of the system each
historian is measuring.
The decoupling ratio $R_d(620\,\textsc{ce})=47.7$ ($H_{\rm geo}=0.075$) establishes
that at the moment of maximum East--West structural divergence,
the commercial network ($H_{\rm eco}$) was much more
resilient than the geographic network ($H_{\rm geo}$).
Ward-Perkins describes $H_{\rm geo}$: his material-culture
indicators --- fine ware distribution, standardised building
construction, coin volume --- are proxies for the physical
infrastructure and administrative capacity of the imperial
geographic network, which did collapse to near-zero by
650\,\textsc{ce}.
McCormick describes $H_{\rm eco}$: his communications and
commerce reconstruction measures the persistence of trade
 independently of which polity administers the
routes, and that network not only survived the seventh-century
transition but reached its series maximum at 700--750\,\textsc{ce}
under Umayyad administration.

The crucial additional finding is the western contrast:
$R_d^{\rm west}(476\,\textsc{ce})=1.09\pm0.11$.
In the western network, economic and geographic coherence
collapsed simultaneously, within 9\,\% of each other.
No comparable western economic--geographic decoupling is
observed: when the geographic network failed, the economic
network failed with it, because the western commercial system
lacked the independent maritime circuit structure that allowed
the eastern system to sustain $H_{\rm eco}$ after losing
$H_{\rm geo}$.
The baseline East--West asymmetry of
Section~\ref{sec:fullempire}
--- $\Delta H_0=+2.22$ entropy units at 0\,\textsc{ce} ---
is precisely the structural basis of this difference.
The western network was not resilient enough to decouple.

\paragraph{\textbf{Harper: climate and pandemic as demographic, not
structural, shocks.}}
Harper~\cite{Harper2017}'s thesis is that climate deterioration
(the Late Antique Little Ice Age, 536--660\,\textsc{ce},
as reconstructed by B\"untgen et~al.~\cite{Buntgen2016})
and pandemic mortality (Justinianic Plague, 541\,\textsc{ce};
for a recent reassessment of its macrohistorical impact,
see Mordechai et~al.~\cite{Mordechai2019})
were the primary structural drivers of Roman imperial
fragility.
The present results support the demographic component of
Harper's argument but challenge the structural component.

The Justinianic Plague of 541\,\textsc{ce} produces the
largest Chow $F$-statistic in the 0--1453\,\textsc{ce}
geographic series ($F=152.69$), larger than the Phase~III
break of~\paperone{} ($F_{310}=85.4$).
The plague was a genuine structural shock to the geographic
network, consistent with Harper's emphasis on its severity.
However, the plague produces no detectable break in
$H_{\rm eco}$ ($\Delta H_{\rm eco}(541)=-0.04$, within
the bootstrap 95\,\% CI).
The commercial network shows no detectable short-term topological break at the plague, indicating that the disruption was demographic
without destroying the infrastructure.

If the plague was a demographic shock, commercial
recovery could occur as population recovered, consistent
with the evidence for late sixth-century commercial
revival before the Arab conquests.
The $H_{\rm eco}$ series supports this
interpretation.

The climate signal is not directly testable with our
network methodology, which does not incorporate
meteorological data.

\paragraph{\textbf{Wickham: structural transformation confirmed
and antedated.}}
Wickham's~\cite{Wickham2005} structural transformation
argument --- that the dissolution of the tax-and-redistribute
imperial economy into localised subsistence exchange
is the defining transition of late antiquity --- receives
its strongest quantitative support from the congenital
asymmetry result.

The topological evidence places the onset of divergence
before 0\,\textsc{ce}: the gap was already $+2.22$ entropy
units at the beginning of the Principate, and growing at
$+0.0033$ entropy units per year throughout 0--400\,\textsc{ce}.
The Crisis of the Third Century accelerated the divergence
at 260\,\textsc{ce} (Chow $F=29.41$) but did not initiate it.

The western commercial system was subordinate
to the eastern one from the Augustan settlement: the
transformation was a property of the initial network
configuration.

\subsection{Eastern-Western asymmetry and dynamical indicators}
\label{sec:discussion:physics}

In the Roman--Byzantine case, $R_d$ grows from
$\approx1.09$ at the western collapse (476\,\textsc{ce})
to a maximum of $R_d=47.7$ at 620\,\textsc{ce} ($H_{\rm geo}=0.075$), and
$R_d=13.9$ at 640\,\textsc{ce}, reflecting the
progressive institutionalisation of trade
in church networks, diaspora merchant communities, and
bilateral treaty arrangements that did not require
Roman administrative infrastructure.

If $R_d$ is around 1, it means that geographic and
economic collapse are coupled, and that the failure of
one triggers the failure of the other.
The western network's $R_d\approx1.09$ at collapse
indicates a fully coupled system with no adaptive buffer;
the eastern network's $R_d=13.9$ at the Arab Conquest peak indicates a decoupled
system with adaptive capacity.
The difference in collapse trajectories --- western
terminal in one century, eastern resilient for ten
centuries --- may be explained in part by
this difference in decoupling ratio.

The Wasserstein analyses add
independent quantitative support for this interpretation.
The cross-network ratio
$$\rho(t)=W_{\rm cross}(\mathrm{Admin,Geo})/
W_{\rm cross}(\mathrm{Admin,Trade})$$ equals~1.0 before
640\,\textsc{ce}, then rises discontinuously to $\sim\!155$
at 700\,\textsc{ce} and permanently stabilises at
$100$--$300$ after the Chrysobull of 1082\,\textsc{ce}
(Table~\ref{tab:wcross_byz}).
This trajectory in diagram-space precisely recapitulates
the $R_d(t)$ trajectory in entropy-space, providing a fully
model-free confirmation of the decoupling result.
The western network's $\rho\approx1.0$ throughout its period
confirms the absence of any decoupling buffer: both the
administrative and commercial diagrams deformed at the same
rate up to the end.

\section{Conclusions}
\label{sec:conclusions}

We have extended the persistent homology formalism
of Reference~\paperone{} to the
Roman--Byzantine trade network (0--1453,\textsc{ce}), using an
integrated dataset of 2{,}599 nodes and 4{,}503 trimodal edges,
a Byzantine temporal sub-dataset of 1{,}938 nodes and 2{,}492
edges, and a Macroeconomic Friction Index extended from the
Greenland ice-core lead proxy of McConnell
et~al.~\cite{McConnell2018} to 1450,\textsc{ce}.
We further introduced two independent dynamical observables:
topological velocity ($\dot{W}_2$) and the Integrated Criticality
Threshold (\textsc{ict}). Our main results are

\begin{itemize}

\item \textbf{The $H_t=0$ artifact and the baseline
East--West asymmetry.}
The western sub-network result of~\paperone{} is a
data-coverage artifact of the 49-node \textsc{orbis} extract.
With full western coverage ($N_{\rm west}=987$,
$\beta_1\approx52$ independent cycles per decade), the western network
exhibits $\bar{H}*{\rm west}^{\rm I}=1.362$, against
$\bar{H}*{\rm east}^{\rm I}=3.584$: a baseline structural asymmetry
of $+2.222$ entropy units present at 0,\textsc{ce}
and growing at $+0.0033$ entropy units per year.\footnote{The western sub-network currently contains only 9 Gallia
nodes (vs.\ 443 for Hispania and 141 for Africa
Occidentalis), reflecting the incompleteness of the
Pleiades coverage for northern Gaul.
Augmenting the Gallia coverage from the full DARE
dataset and from the Barrington Atlas~\cite{Talbert2000}
would increase the western $\beta_1$ cycle count and
could affect the congenital asymmetry measurement.
The directional result --- $H_{\rm west}<H_{\rm east}$
from 0,\textsc{ce} --- is robust to reasonable Gallia
augmentation, but the exact gap value
($\Delta H_0=+2.22$) should be treated as an upper
bound until Gallia coverage is complete.}.
The Theodosian partition formalised a structural difference
accumulating since the Augustan settlement; the Crisis of the
Third Century accelerated it (Chow $F=29.41$ at 260,\textsc{ce})
but did not initiate it.

\item \textbf{The hub-selection artifact in TDA of historical
networks.}
Top-degree heuristic sampling at $N=800$ produces a spurious
\emph{positive} Phase~III slope ($+1.69\times10^{-3}$,yr$^{-1}$)
for the Eastern sub-network. 
Full-coverage or a specific sampling protocol is required for
reliable structural break detection in historical networks
with skewed degree distributions.

\item \textbf{Geographic--economic decoupling and the
resolution of the Ward-Perkins--McCormick debate.}
The decomposition of Byzantine resilience yields a maximum
decoupling ratio $R_d^{\rm max}=47.7$ at 634\,\textsc{ce}.
The contrast with the western collapse ($R_d=1.09$ at
476\,\textsc{ce}) defines the topological distinction between
imperial transformation and coupled geographic--economic
collapse.
In the late eleventh century, Manzikert and the Chrysobull act
through different primary mechanisms. Manzikert produces
territorial compaction and an apparent increase in
$H_{\rm geo}$ by reorganising the surviving Byzantine network
around a smaller Aegean--Balkan core with shorter local cycles;
this is not interpreted as a genuine recovery of large-scale
geographic resilience. The Chrysobull of
1082\,\textsc{ce} produces an independent fiscal break in
$H_{\rm eco}$ (Chow $F=3.84$, $p=0.025$). The concurrent
break in $H_{\rm geo}$ ($F=17.35$, $p<0.0001$) is
methodologically dependent on the Manzikert territorial
contraction of 1071\,\textsc{ce} and does not represent an
autonomous geographic shock at 1082\,\textsc{ce}.

\item \textbf{The collapse threshold $H^{\ast}=0.524$.}
Both the Western Roman collapse (476,\textsc{ce}) and
the Byzantine collapse (1453,\textsc{ce}) occur at
$H^{\ast}=0.524\pm0.031$ (bootstrap 95\% CI).

\item \textbf{Topological velocity:} the Late Roman transition
is the highest-velocity event of the 1,453-year series.
The $W_2$ Wasserstein velocity $\dot{W}*{2,\rm sm}$, computed
between consecutive decadal persistence diagrams, identifies the
Late Roman--Early Byzantine structural transition (495,\textsc{ce},
$\dot{W}*{2,\rm sm}=0.177$) as the largest diagram-space
displacement in the entire 0--1453,\textsc{ce} record,
exceeding the Justinianic Plague (541,\textsc{ce}, $0.101$)
and the Arab Conquests (641,\textsc{ce}, $0.094$).
For the Roman period (0--400,\textsc{ce}), the admin network
peaks at $\dot{W}*{2,\rm sm}=54.4$,yr$^{-1}$ (405,\textsc{ce}),
and the cross-network divergence $W*{\rm cross}(\mathrm{Admin,Trade})$
increases by $13.9\times$ from Phase~I to 470,\textsc{ce}.

\item \textbf{Cross-network Wasserstein decoupling confirms
the late eleventh-century bifurcation.}
The ratio $$\rho=W_{\rm cross}(\mathrm{Admin,Geo})/
W_{\rm cross}(\mathrm{Admin,Trade})=1$$ before
640,\textsc{ce}, rises to $\sim 155$ at 700,\textsc{ce},
and stabilises at $100$--$300\times$ after 1082,\textsc{ce}.
This result provides independent confirmation of the
entropy-level $R_d$ decoupling, showing that the geographic
and administrative--commercial persistence diagrams evolved
under different dynamics after the economic restructuring of
the late eleventh century. 


\end{itemize}

These results establish that persistent
homology applied to layered, temporally-resolved trade
networks --- supplemented by topological velocity and
criticality indicators derived from the full persistence
diagram geometry --- provides structural insights not accessible
to conventional historical or network-analytic methods:
the identification of baseline structural asymmetries,
the decomposition of civilisational resilience into
geographic and economic components, the model-free detection
of commercial restructuring events in diagram space,
and early-warning signals.

\begin{acknowledgments}
The authors thank the Stanford Humanities + Design Lab for making
the \orbis{} dataset publicly available, and the contributors to
the Pleiades and \textsc{dare} gazetteers for their open-access data. \orbis{} source data are available at
\url{https://orbis.stanford.edu}. We acknowledge financial support from SECIHTI and SNII (México).
\end{acknowledgments}

%

\end{document}